\documentclass[leqno,12pt]{article}
\usepackage[hang]{subfigure}
\usepackage{color}
\usepackage{latexsym}
\renewcommand{\epsilon}{\varepsilon}

\newcommand{\E}{\mbox{I\negthinspace E}}

\newcommand{\X}{\textbf{X}}
\newcommand{\x}{\textbf{x}}
\newcommand{\param}{\boldsymbol{\theta}}
\newcommand{\argmax}{\textrm{argmax}}
\newcommand{\argmin}{\textrm{argmin}}

\usepackage{amsmath}
\usepackage{amssymb}
\usepackage{ntheorem}
\usepackage[ansinew]{inputenc}
\usepackage{graphicx}
\usepackage{epsfig}
\usepackage{dsfont}

\usepackage[authoryear]{natbib}

\newtheorem{satz}{Theorem}[section]

\newtheorem{rem}[satz]{Remark}

\def\3{\ss}

\newcommand{\bea}{\begin{eqnarray*}}
\newcommand{\eea}{\end{eqnarray*}}
\newcommand{\be}{\begin{eqnarray}}
\newcommand{\ee}{\end{eqnarray}}

\newcommand{\ba}{\begin{array}}
\newcommand{\ea}{\end{array}}
\def\3{\ss}

\begin{document}

\title{Misspecification in copula-based regression}

\author{ Holger Dette, Ria Van Hecke, Stanislav Volgushev \\
Ruhr-Universit\"at Bochum \\
Fakult\"at f\"ur Mathematik \\
44780 Bochum \\
Germany \\
}

 \maketitle

\begin{abstract}
In a recent paper \cite{noh13} proposed a new semiparametric estimate of a regression function with a
multivariate predictor, which is based on a  specification of the dependence structure between the predictor and the response
by means of a parametric copula. This paper investigates the effect which occurs under misspecification of the parametric model.
We demonstrate that even for a one  or two dimensional predictor the error caused by a ``wrong'' specification of the parametric
family is rather severe, if the regression is not monotone in one of the components of the predictor. Moreover, we also show that these
problems occur for all of the commonly used copula families and we illustrate in several examples that the copula-based regression
may lead to invalid results even when more flexible copula models such as vine copulae (with the common parametric families) are used in the estimation procedure.

\end{abstract}

Keywords: curse of dimensionality, semiparametric inference, copulae,
pairwise copulae, vine copulae

\section{Introduction}\label{sec1}
\def\theequation{1.\arabic{equation}}
\setcounter{equation}{0}

Nonparametric regression has emerged as a rather flexible tool to describe the relation between a response $Y$ and a predictor $\textbf{X}$. Since the pioneering work of Nadaraya and Watson [\cite{nadaraya1964}  and \cite{watson1964}] numerous authors have worked on this problem and various estimates such as smoothing splines, local polynomials or series estimators have been proposed and investigated in the literature [see for example \cite{fangij1996}]. It is well known that nonparametric regression estimates suffer from the curse of dimensionality if the dimension of the predictor is large. In this case a regression function cannot be estimated with reasonable accuracy and several authors have proposed methods to avoid this problem. A common feature of all publications in this direction consists in additional structural or parametric assumptions regarding the unknown regression function, such as additivity [see \cite{stone1985}], tree-based models [\cite{hastibfri2001}] or single index models [\cite{ichimura1993}]. In a recent paper \cite{noh13} introduced a novel semiparametric estimate of the regression function in a nonparametric regression model with a high-dimensional predictor. Roughly speaking, these authors propose to model the dependency structure between the response and the predictor by a parametric copula family in order to obtain estimates of the regression function which converge with a parametric rate. The authors demonstrate (theoretically and empirically) that the resulting estimates have nice properties if the parametric copula family has been chosen correctly. \\
This note is devoted to a careful investigation of the properties of the copula-based regression estimate proposed by \cite{noh13} in the case where the copula family is misspecified. {More precisely, our aim is to investigate the kinds of regression dependence that can be described by commonly used copula models. See Section \ref{sec3} for a precise list of copula families that we considered.}
For one-dimensional predictors it will be demonstrated that copula-based regression estimation yields satisfactory results for monotone regression functions if the copula is chosen appropriately. On the other hand, if the regression function is not monotone, copula-based regression estimates do not reproduce the qualitative features of the regression function. This property does not depend on the specific misspecified copula model but can be observed for all of the commonly used parametric copula families. For a high-dimensional predictor the situation is even worse. Moreover, we also show that more flexible models as vine copulae (based on the commonly used parametric models) will not improve the properties of the estimator. The reason for these observations is simple: all of the commonly available parametric copula models are not flexible enough to describe a non-monotone behaviour between the response and one of the predictors.\\
The remaining part of this note is organized as follows. In Section \ref{sec2} we introduce some notation and recall the estimate proposed by \cite{noh13}. Section \ref{sec3} is devoted to some illustrative examples in the one  and two-dimensional case, while some conclusions and additional explanations are given in Section \ref{sec4}.

\section{Copula-based regression inference} \label{sec2}
\def\theequation{2.\arabic{equation}}
\setcounter{equation}{0}
In this section we briefly review the copula-based estimation of a regression function as introduced by \cite{noh13}. To be precise let
 $Y$   and $\textbf{X}=(X_1,\dots,X_d)^T$ be  a real and  $d$-dimensional random  variable ($d\geq 1$), respectively,  and
 denote by  $F_Y,F_1,\dots,F_d$  the cumulative distribution functions of $Y$ and the margins of $\X$, which will be assumed
 as differentiable throughout this note.  The corresponding densities are denoted by $f_Y,f_1,\dots,f_d$. The famous
Sklar's theorem [\cite{sklar59}] shows that the joint distribution function $F$ of the vector $(Y,\X^T)^T$ can be represented as
\[
F(y,\x^T)=C(F_Y(y),F_1(x_1),\dots,F_d(x_d)),
\]
where  $(y,\x^T)^T=(y,x_1,\ldots , x_d)^T$ and $C$ is the copula. \cite{noh13} showed that the mean regression function
\[
m(x_1,\dots,x_d)=\E[Y | \X=(x_1,\dots,x_d)]
\]
%
%
can be represented as
\begin{equation} \label{mcop}
m(x_1,\dots,x_d)=\int_{-\infty}^{\infty} y\frac{c(F_Y(y),F_1(x_1),\dots,F_d(x_d))}{c_{\X}(F_1(x_1),\dots,F_d(x_d))}dF_Y(y) ,
\end{equation}
where $c$ denotes the density of the copula $C$, that is 
$$c= {\partial^{d+1} \over \partial y \partial x_1 \ldots \partial x_d} C$$ and
\[
c_{\X}(u_1,\dots,u_d)=\int_{- \infty}^{\infty} c(F_Y(y),u_1,\dots,u_d)dF_Y(y),
\]
denote the copula densities  corresponding to the distribution of the vectors $(Y,\X^T)^T$ and  $\X$, respectively. In order to avoid the
curse of dimensionality in the estimation of the regression function $m$ these authors  propose to use a semi-parametric estimate using a parametric copula family, say $\{c_\theta |~\theta \in \Theta \}$
for the copula density $c$ in \eqref{mcop} and to estimate the unknown marginal distributions separately. More precisely, if $(Y_1,\X_1^T)^T, \ldots , (Y_n,\X_n^T)^T$ denotes a sample of independent identically distributed observations with copula $C$ and marginal distribution functions $F_Y, F_1, ... , F_d$,
\cite{noh13} suggest  to estimate the marginal distributions
$F_Y$ and $F_j$ nonparametrically by
\[
\hat{F}_Y(y)=\frac{1}{n+1}\sum_{i=1}^{n} I(Y_i \leq y)
\mbox{~~ and  }
\hat{F}_j(x)=\frac{1}{n+1}\sum_{i=1}^{n} I(X_{ij} \leq x),
\]
respectively (here we use the notation $\X_i=(X_{i1},\dots,X_{id})^T$)  and to estimate the parameter $\param$ of the parametric copula family
by a pseudo--maximum likelihood method, that is
\[
\hat{\param}_{PL}=\underset{\param \in \Theta}{\argmax}\sum_{i=1}^n\log c (\hat{F}_Y(Y_i),\hat{F}_1(X_{i1}),\dots,\hat{F}_d(X_{id});\param )
\]
[see \cite{genghoriv1995} or \cite{tsukahara2005}]. The final estimate of the regression function  $m$  is then defined by
\begin{eqnarray} \label{estim}
\hat{m}(\x)&=&\int_{-\infty}^{\infty} y \frac{{c}(\hat{F}_Y(y),\hat{F}_1(x_1),\dots,\hat{F}_d(x_d);\hat{\param}_{PL})}{\int_{-\infty}^{\infty}  {c}(\hat{F}_Y(u),\hat{F}_1(x_1),\dots,\hat{F}_d(x_d);\hat{\param}_{PL})d\hat{F}_Y(u) } d\hat{F}_Y(y) \\
&=& \frac{1}{n}\sum_{i=1}^nY_i\frac{c(\hat{F}_Y(Y_i),\hat{F}_1(x_1),\dots,\hat{F}_d(x_d);\hat{\param}_{PL})}{\frac{1}{n}\sum_{j=1}^nc(\hat{F}_Y(Y_j),\hat{F}_1(x_1),\dots,\hat{F}_d(x_d);\hat{\param}_{PL})}
\nonumber
\end{eqnarray}
In the case of a one-dimensional  covariate, i.e. $d=1$, we have  $c_{X_1}\equiv 1$ and thus the estimate simplifies to
\begin{eqnarray} \label{estim1}
\hat{m}(x)=\frac{1}{n}\sum_{i=1}^nY_ic(\hat{F}_Y(Y_i),\hat{F}_1(x);\hat{\param}_{PL}).
\end{eqnarray}
\cite{noh13} demonstrate that the estimator defined in \eqref{estim} avoids the problem of the curse of dimensionality. More precisely, they show that $\hat{m}(\x)$ is a $\sqrt{n}$-consistent and asymptotically normal distributed estimate if the parametric copula model has been specified correctly. On the other hand, under misspecification of the copula structure it is shown that the statistic $\hat{m}(\x)$ estimates the quantity
\begin{equation} \label{best}
m({\bf x}, \param^*) = \int_{- \infty}^{\infty} y \frac {c(F_Y (y), F_1(x_1), \ldots , F_d(x_d), \param^*)} {c_{\X} (F_1(x_1), \ldots , F_d(x_d), \param^*)} d F_Y(y) ,
   \end{equation}
   where $\theta^*$ is the minimum of the function
   \begin{equation}
   I(\param) = \int_{[0,1]^{d+1}} \log \Big( \frac {c(u_0, \ldots , u_d)} {c(u_0, \ldots , u_d, \param)} \Big) d C(u_0, \ldots , u_d).
   \end{equation}
{As it was  pointed out by \cite{noh13}, 
   the quantity $m({\bf x}, \param^*)$ does  in general  not coincide with the true regression function $m({\bf x})$.
Consequently there exists a bias if the parametric copula has been misspecified, 
    but no further evidence regarding the kinds of regression functions which can be estimated well 
    (i.e. for which this bias is small) is given. Overall, one might hope that the commonly used parametric copula models are flexible enough to model a rich variety of regression dependencies.}
	{In the following section however we will demonstrate that this is not the case and that the quality of the estimate \eqref{estim} under misspecification of the parametric copula depends heavily on the specific structure of the unknown regression function $m$.} In particular we show that for non-monotone regression functions these estimates are in fact not reliable. We will also demonstrate that model selection from a class of the commonly used copula families by information type criteria (in the case $d=1$) or the application of more flexible copula families such as vine copulae in the case $d\geq 2$ [see \cite{bedcoo2002,aas09}] does not solve these problems. As soon as  the regression is not monotone in one of the components of the explanatory variable the copula based regression estimate and the true regression function show substantially different qualitative  features.

\section{Inference under misspecification - examples} \label{sec3}
\def\theequation{3.\arabic{equation}}
\setcounter{equation}{0}

In this section we provide several examples indicating some difficulties in copula based regression inference as proposed by \cite{noh13}. All presented results are based on a sample of size $n=100$
and for the simulation of the mean squared error we use $1000$ simulation runs. For the sake of brevity we restrict ourselves to
 the case $d=1$ and $d=2$, where the problems of misspecification of the parametric copula family are already very visible and the
 arguments are more transparent. We expect that for high dimensional predictors these problems are even more severe.

\subsection{One-dimensional predictors}

\begin{figure}[t]
\includegraphics[height=6cm,width=6cm]{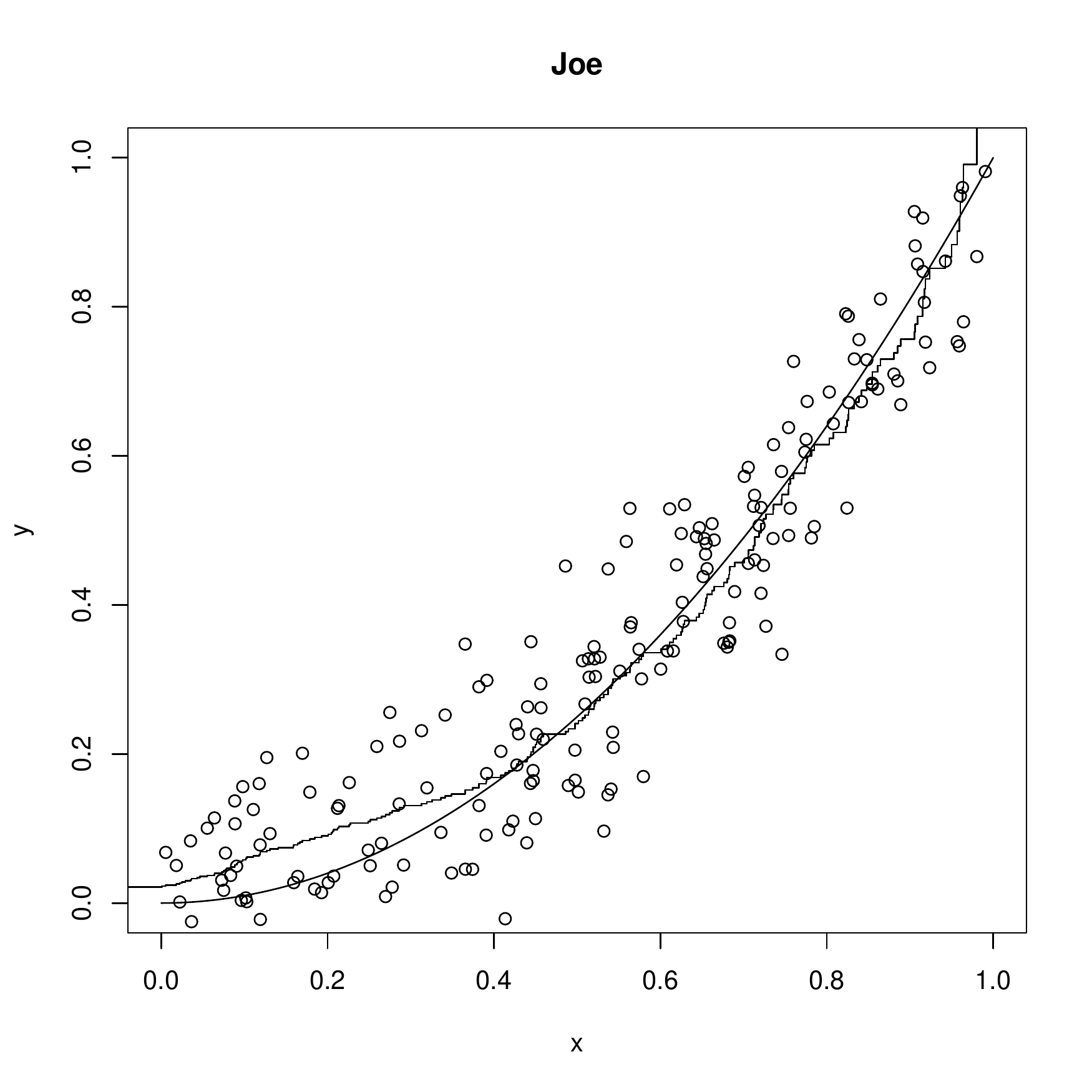}
\includegraphics[height=6cm,width=6cm]{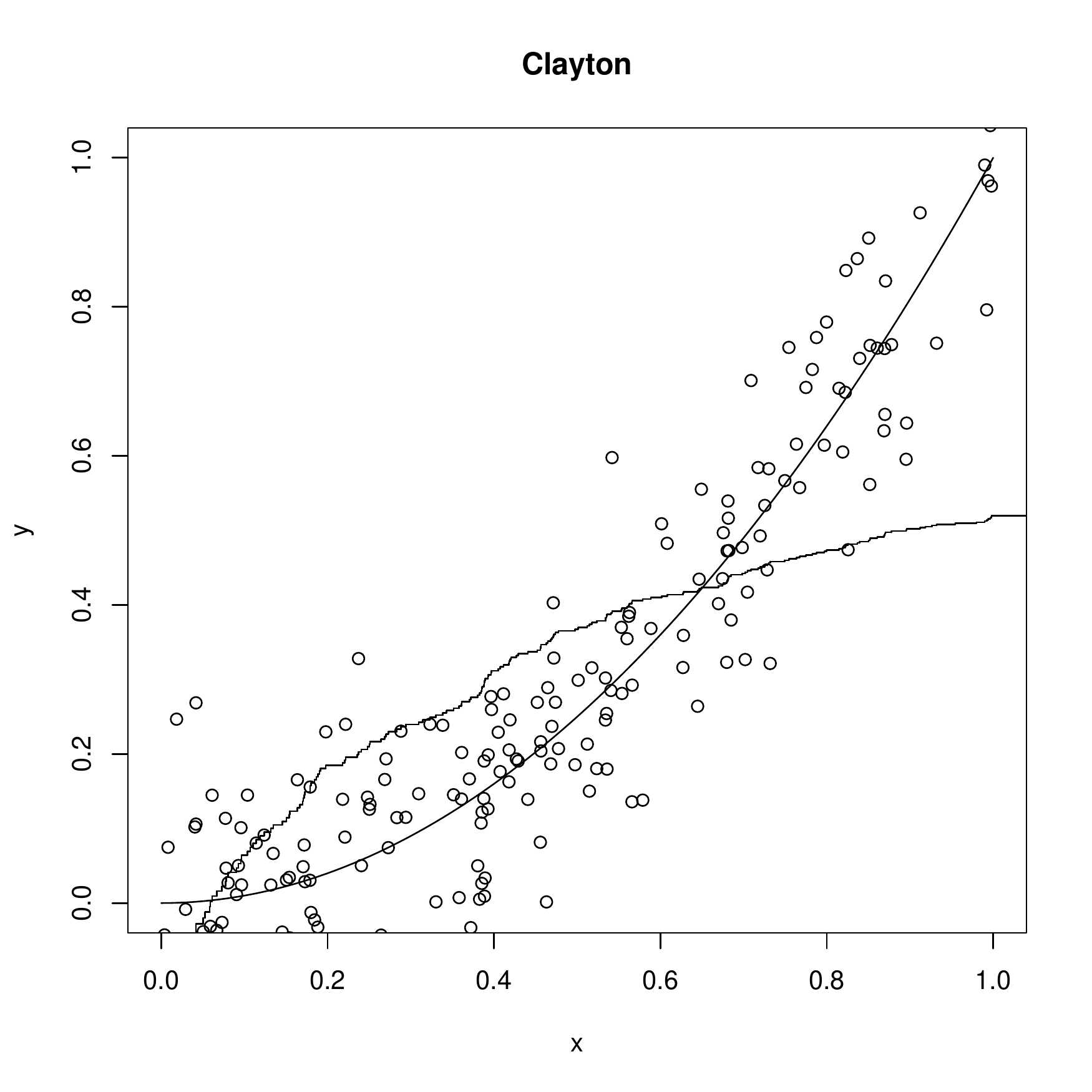}
\includegraphics[height=6cm,width=6cm]{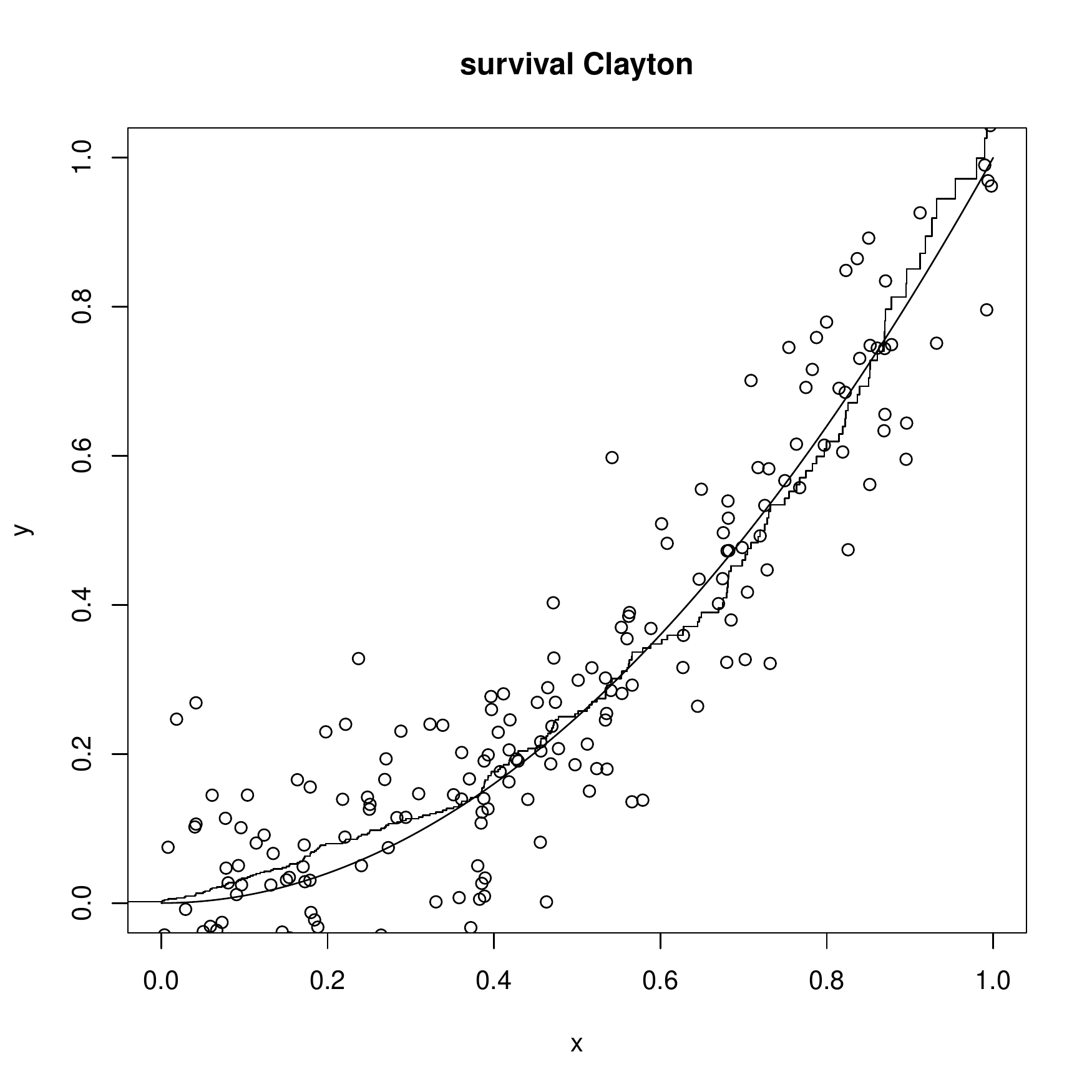}
\caption{\label{fig1}{\it Copula based regression estimates of the regression model \eqref{m1}. The Joe copula (left panel), Clayton copula (middle panel) and  survival Clayton copula
(right panel) are used in the estimate \eqref{estim}.}}
\end{figure}

\begin{figure}[t]
\includegraphics[height=6cm,width=6cm]{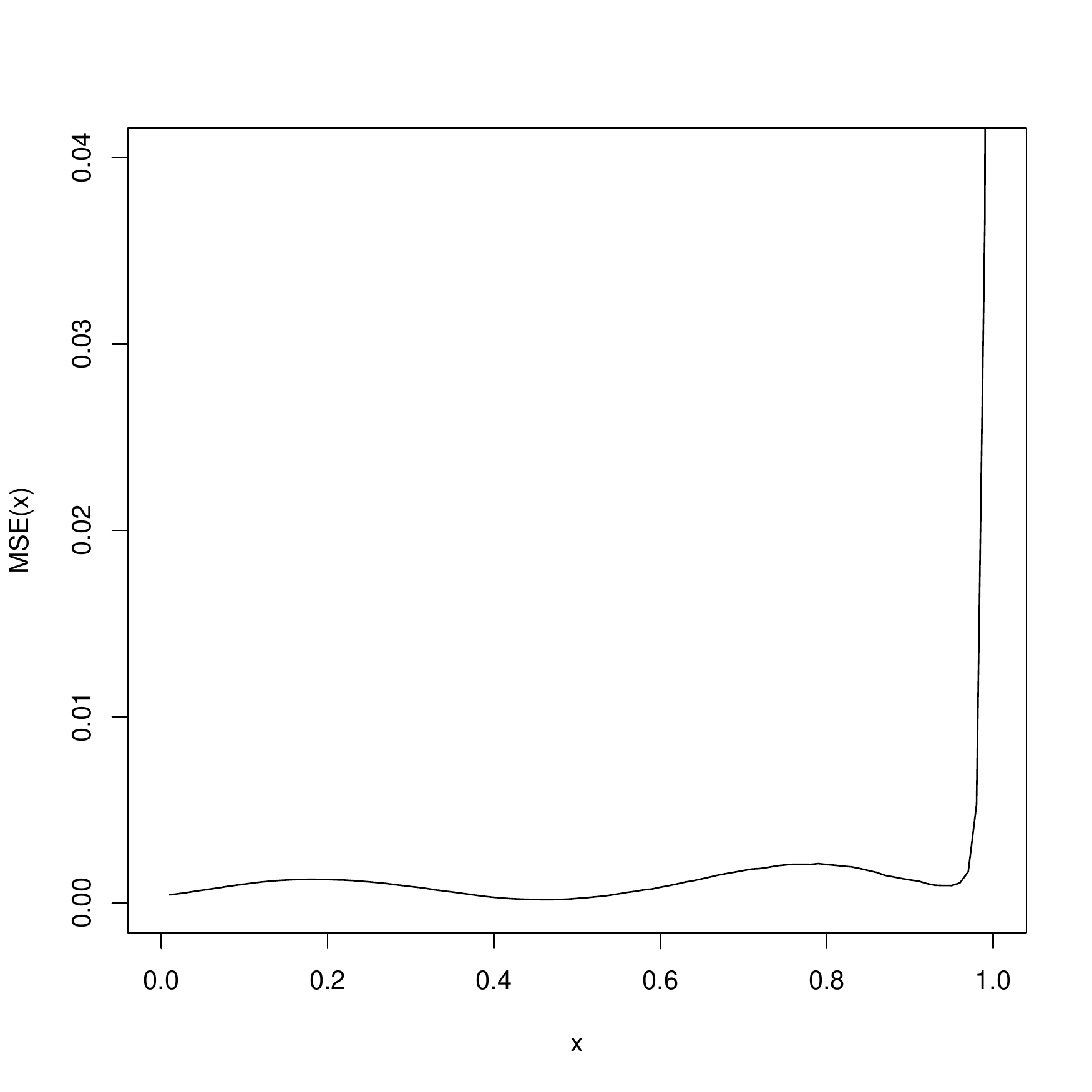}
\includegraphics[height=6cm,width=6cm]{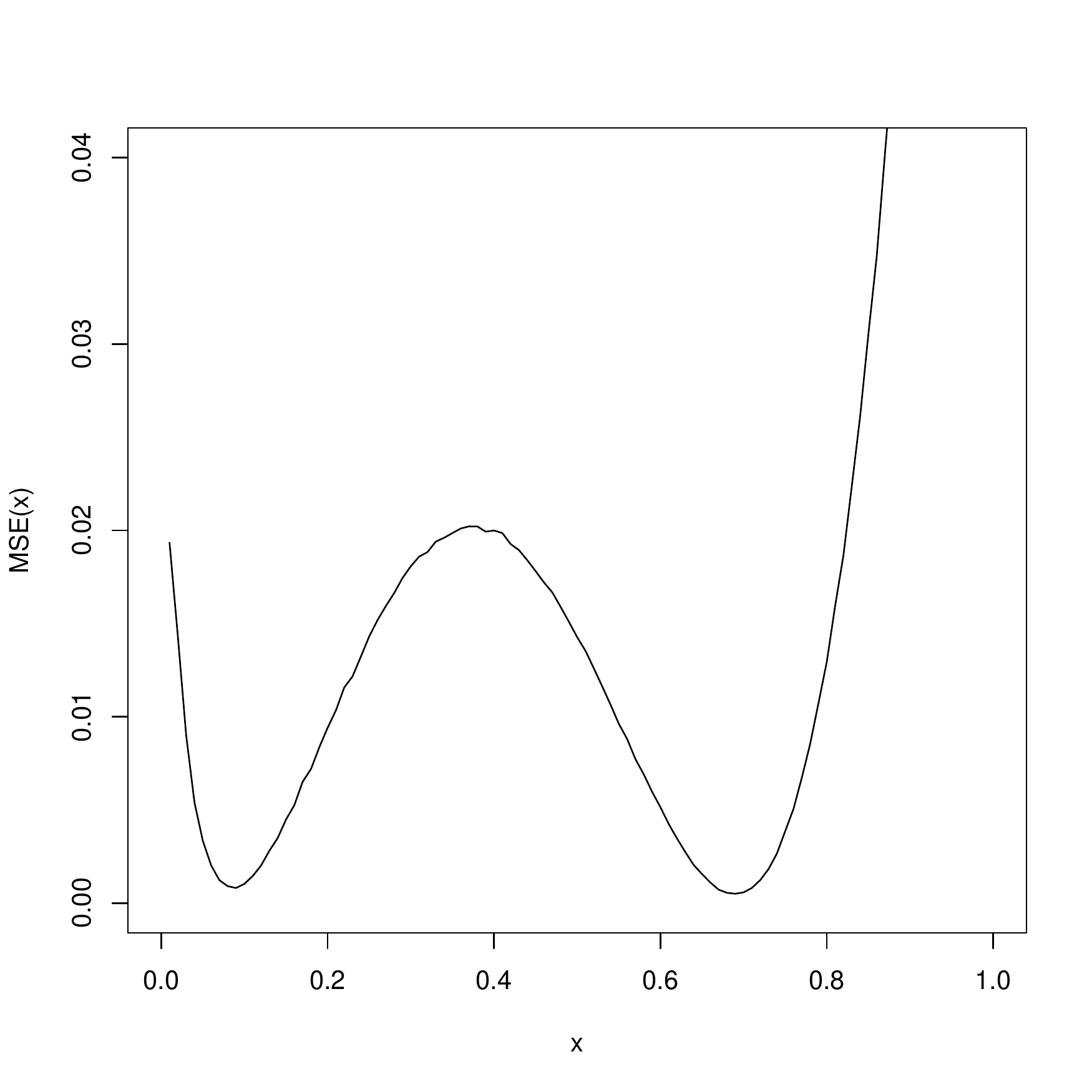}
\includegraphics[height=6cm,width=6cm]{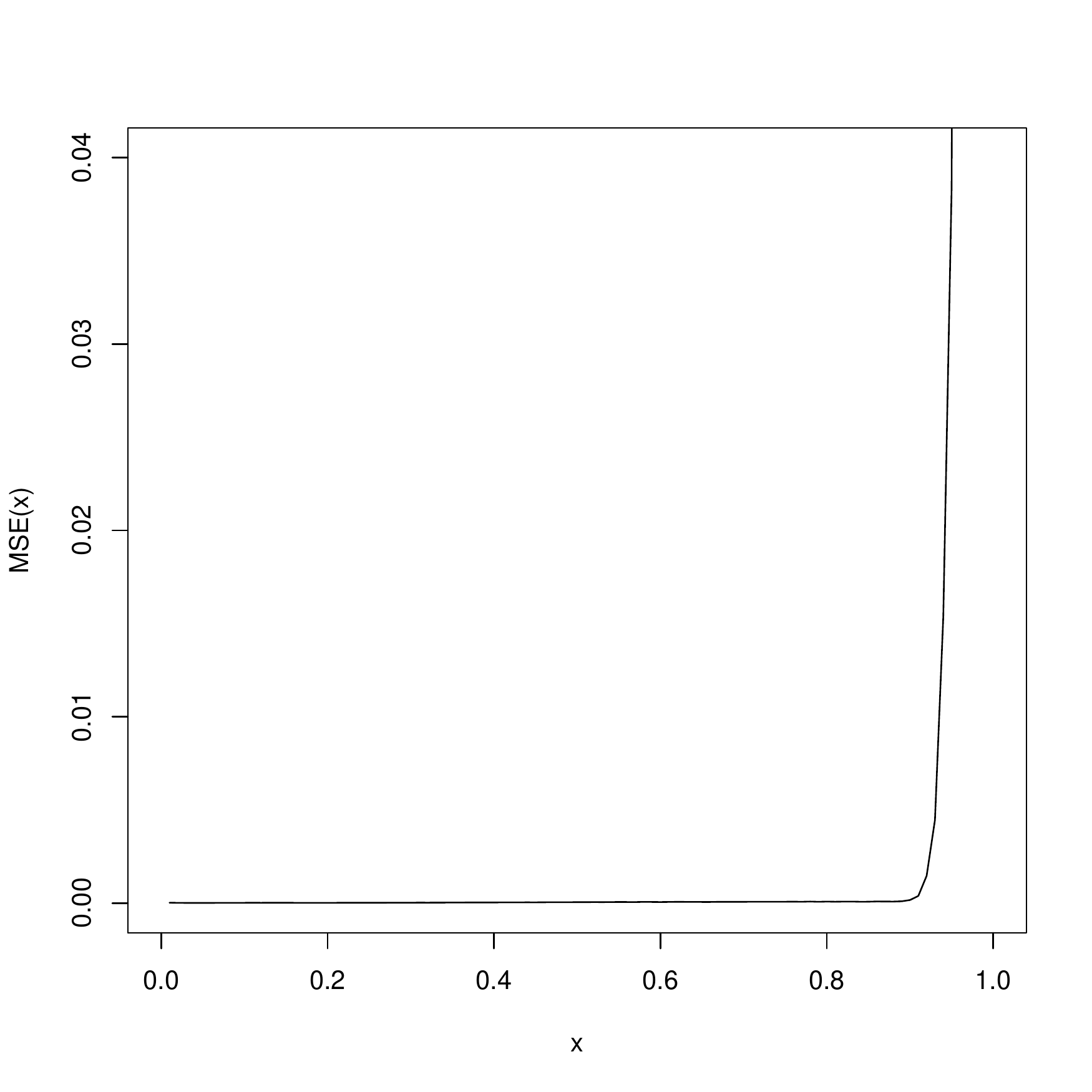}
\caption{\label{fig2}{\it Mean squared error of the  copula based regression estimates of the regression function in model \eqref{m1}. The Joe copula (left panel), Clayton copula (middle panel) and survival Clayton copula
(right panel) are used in the estimate \eqref{estim}.}}
\end{figure}

We start our investigation with the one-dimensional nonparametric regression model
\begin{equation} \label{reg}
Y_i=m(X_i) +\sigma \varepsilon_i,~~i=1,\ldots , n,
\end{equation}
where the
explanatory variable $X_i$ in the regression model \eqref{reg}  is uniformly distributed on the interval $[0,1]$ and
the errors are normally distributed with mean $0$ and variance $\sigma^2=0.01$.  The regression functions are given by
\begin{eqnarray}  \label{m1}
m(x_1)&= &x_1^2 \\
m(x_1)&= &(x_1-0.5)^2
\label{m2}
\end{eqnarray}
and reflect a typical increasing and convex case. Exemplarily we present in Figure \ref{fig1}  "typical" simulated data from these models
with the corresponding copula regression estimates. We have investigated most of the commonly used copula models, but for the sake of brevity we show only results for three cases, namely the Joe, Clayton and the survival Clayton copula [see \cite{nelson2006}]. The application of the Joe copula yields a rather reasonable fit, while the Clayton copula does not show satisfactory results. On the other hand the application of the survival Clayton copula again produces a reasonable fit. These findings are also reflected in the mean squared error curves, which are shown in Figure \ref{fig2} and we conclude that some care is necessary in the choice of the parametric copula family for the estimate (\ref{estim}), even if the regression function is monotone. If this family has been chosen appropriately, and the regression function is monotone, the statistic \eqref{estim} produces reliable estimates of the regression function $m$ with a small bias.

\begin{figure}[t]
\includegraphics[height=6cm,width=6cm]{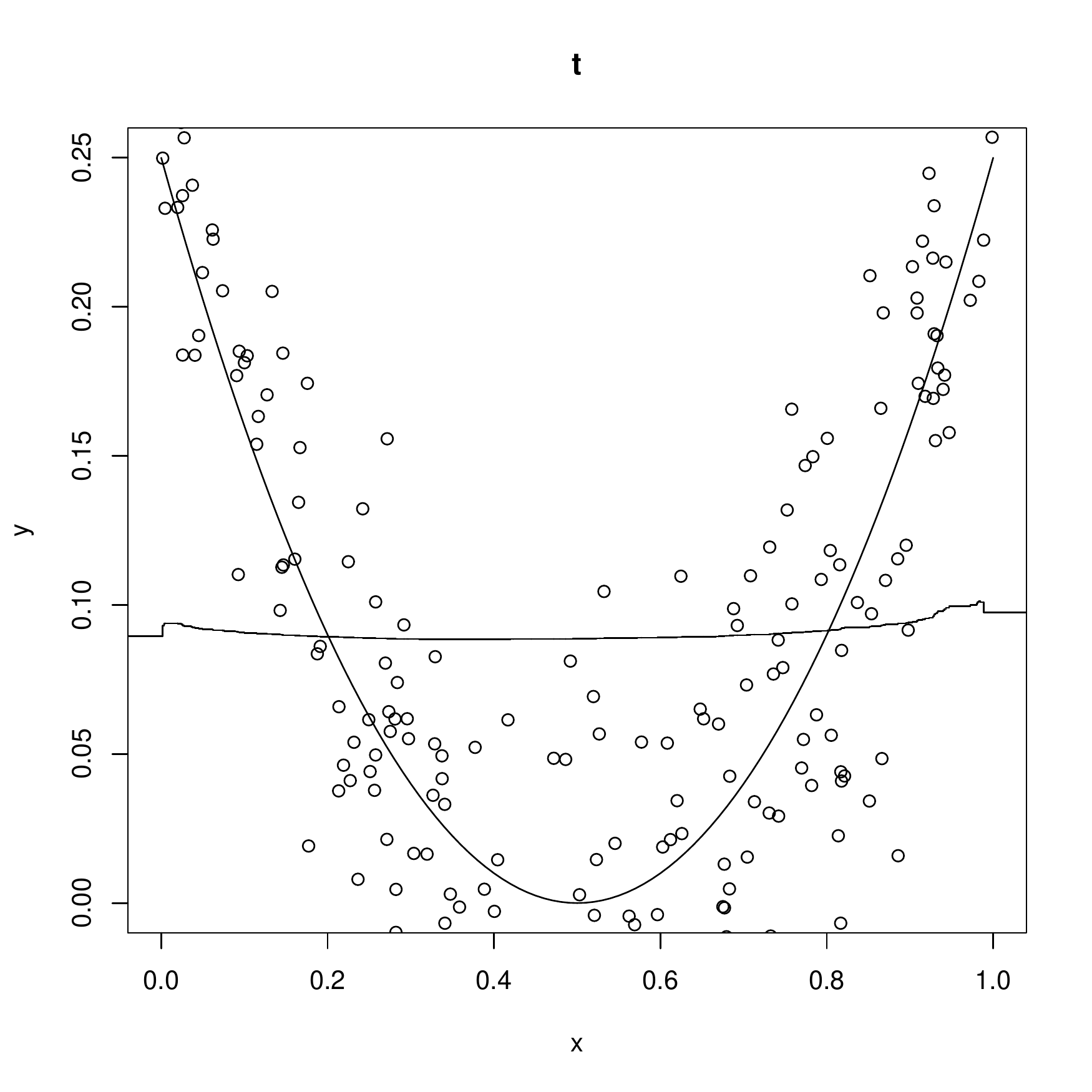}
\includegraphics[height=6cm,width=6cm]{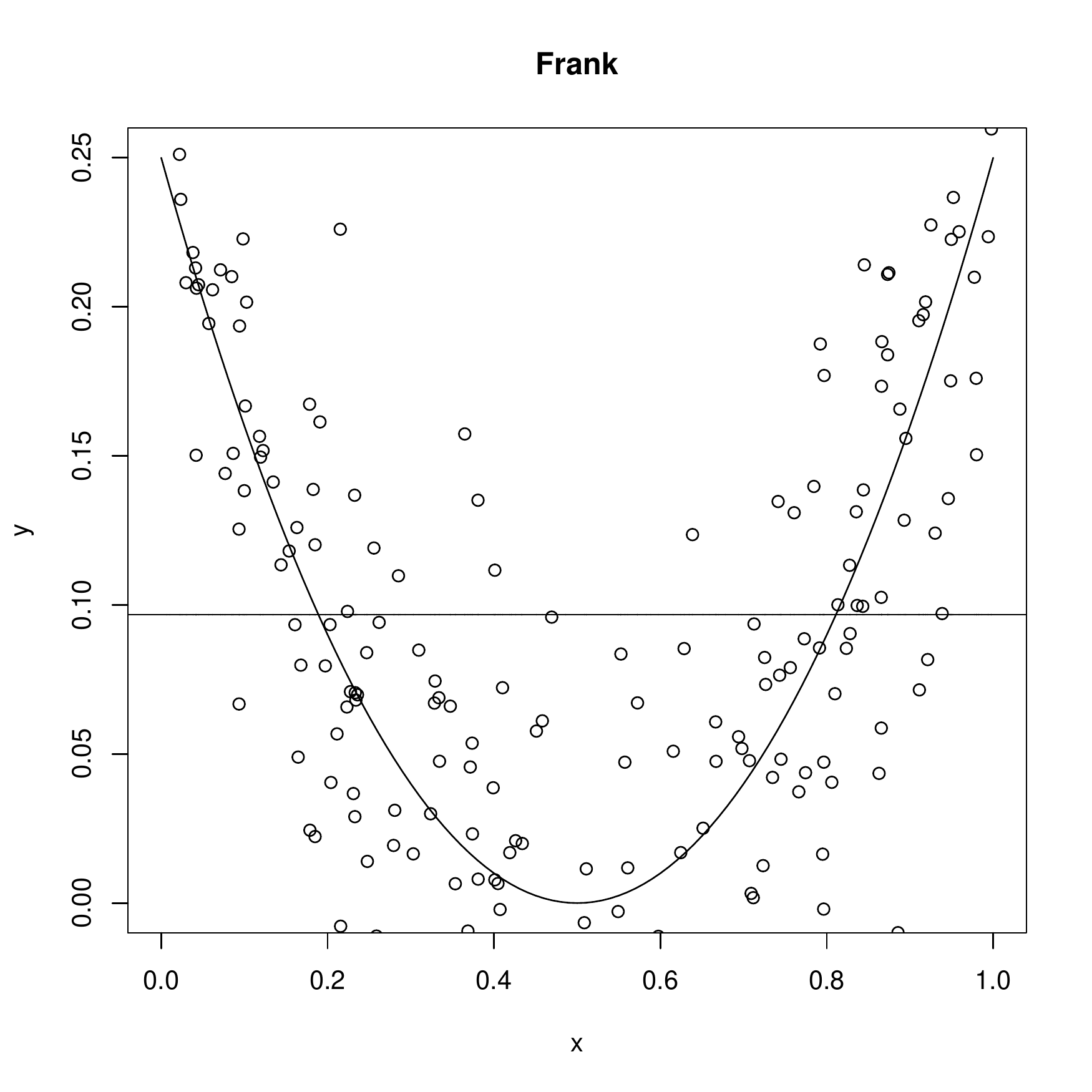}
\includegraphics[height=6cm,width=6cm]{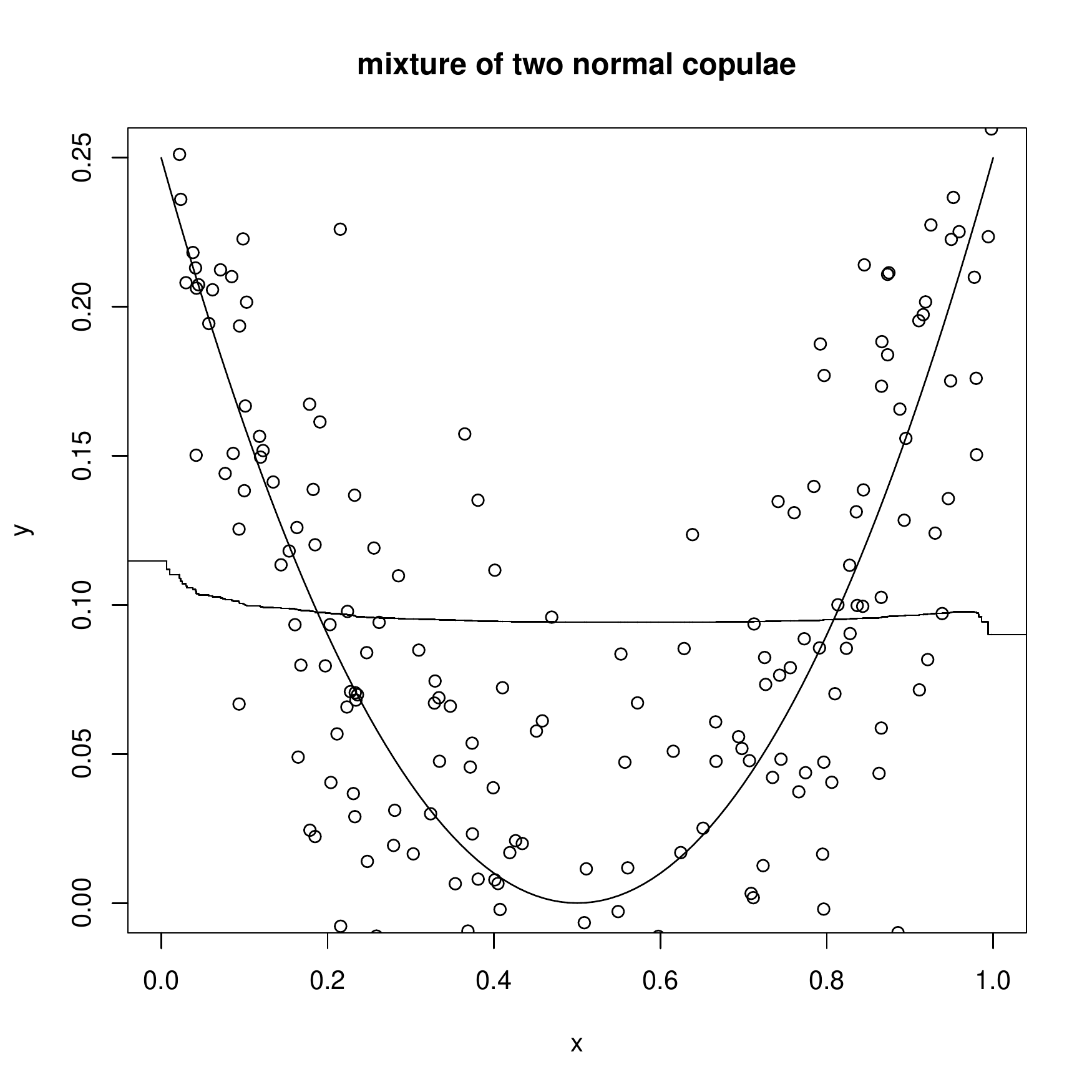}
\caption{\label{fig3}{\it Copula based regression estimates of the regression function in model \eqref{m2}. The student copula (left panel), Frank copula (middle panel) and a mixtures of two normal copulae (right panel) are used in the estimate \eqref{estim}.}}
\end{figure}

\begin{figure}[t]
\includegraphics[height=6cm,width=6cm]{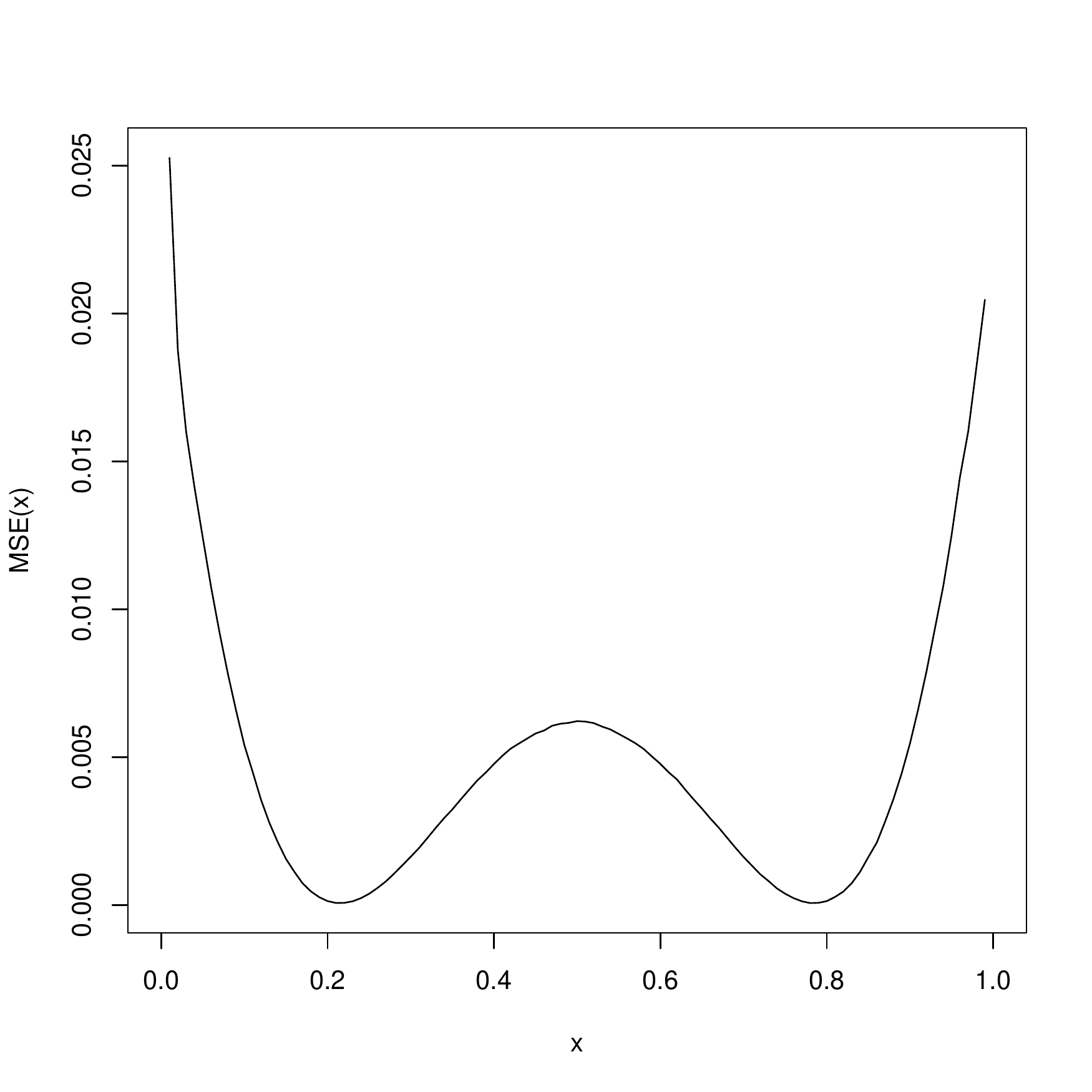}
\includegraphics[height=6cm,width=6cm]{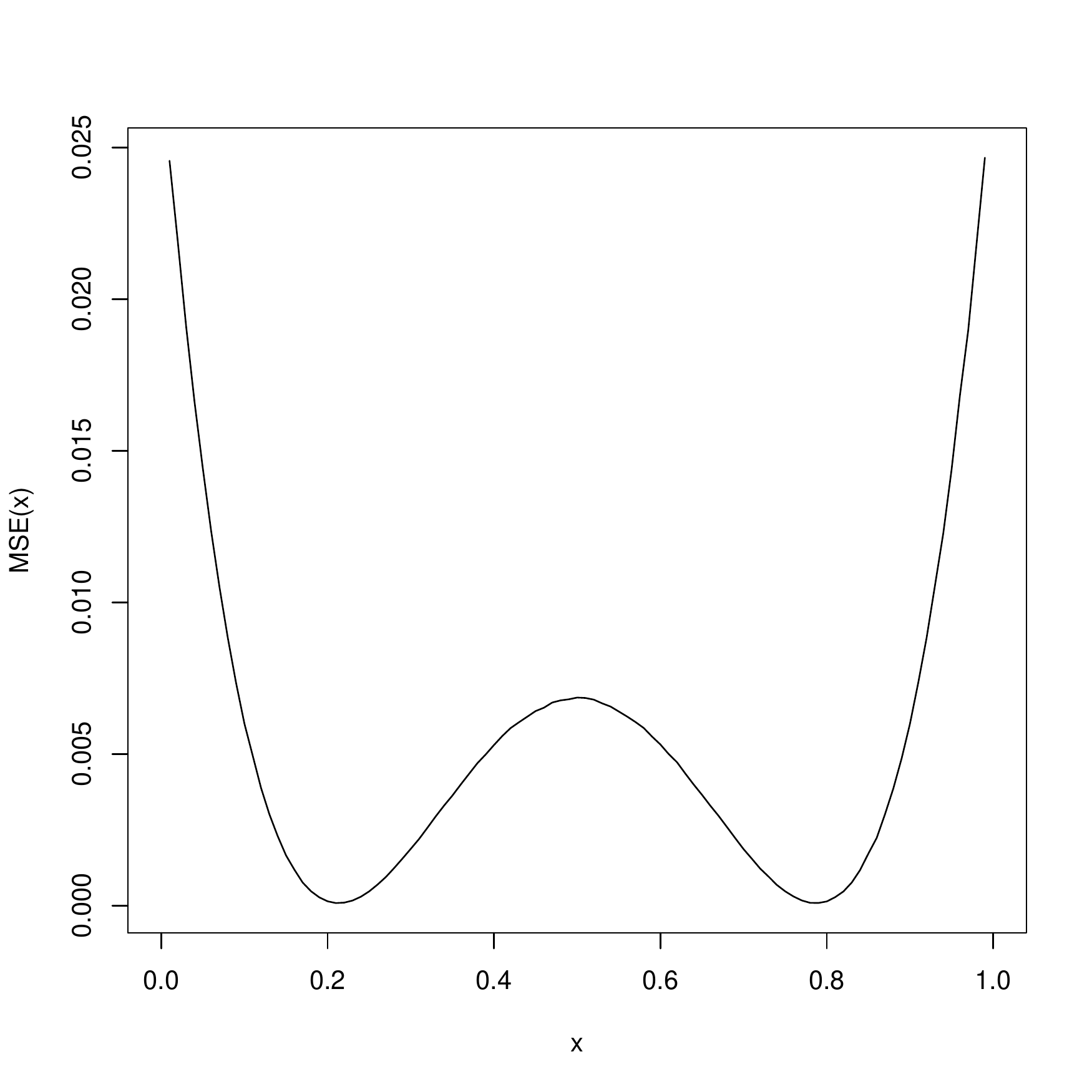}
\includegraphics[height=6cm,width=6cm]{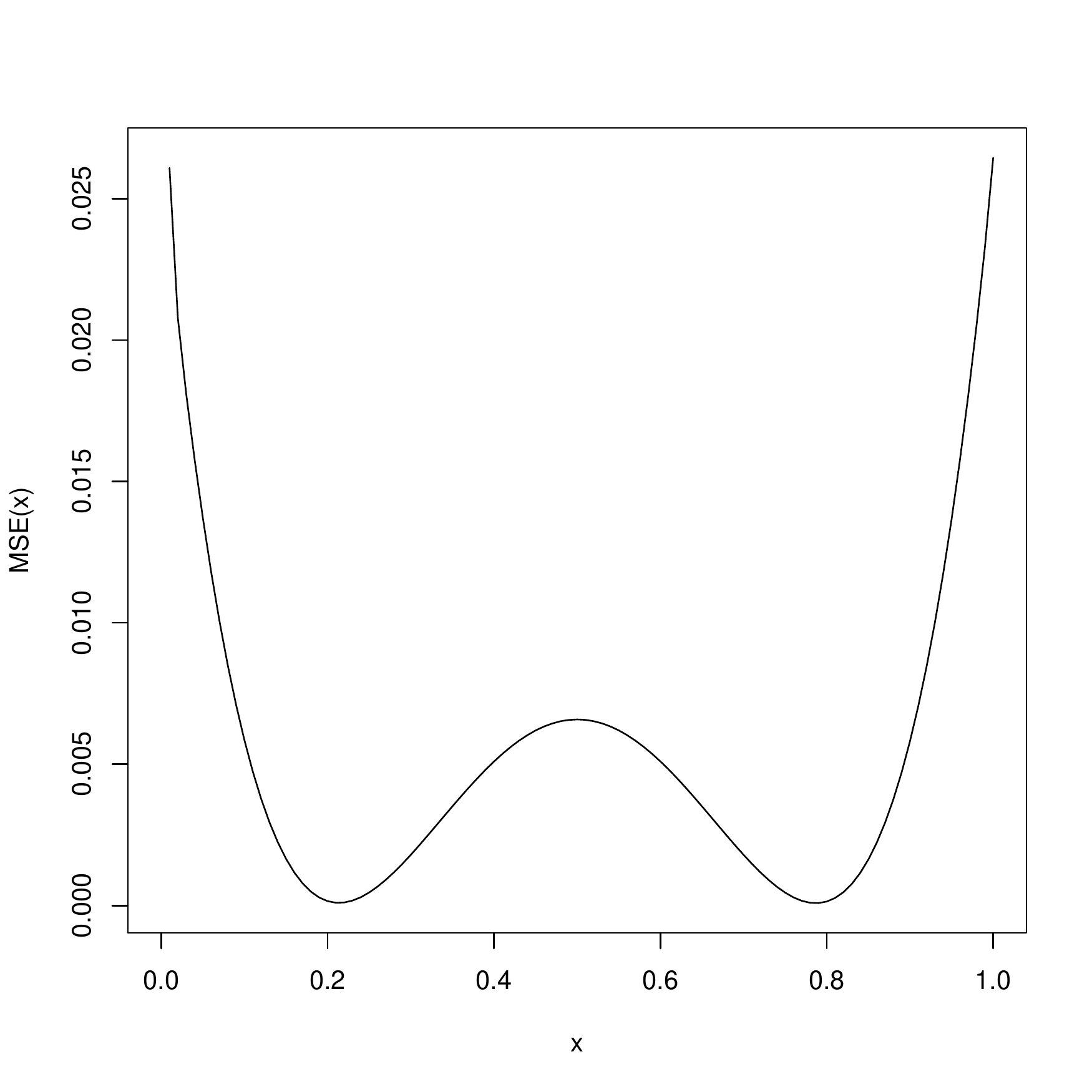}
\caption{\label{fig4}{\it Mean squared error of the  copula based regression estimates of the regression function in model \eqref{m2}. The student copula (left panel), Frank copula (middle panel)
and a mixtures of two normal copulae (right panel) are used in the estimate \eqref{estim}.}}
\end{figure}

On the other hand, if the regression function is not monotone, the situation is completely different, which is demonstrated by our second example considering the regression function \eqref{m2}. Corresponding results are shown in Figure \ref{fig3} for typical simulated data, where we use the $t$, Frank copula in the left and middle panel
and a mixture of two normal copulas in the right panel (here the mixing proportion is also estimated from the data). The corresponding mean squared errors are depicted in Figure \ref{fig4} and as a first conclusion we note that none of these choices yields a reasonable estimate of the regression function. {In fact we considered all copulae from the following list \{``amh copula'',``independence copula'', ``Gaussian copula'', "t-copula", ``Clayton copula'', ``Gumbel copula'',
  ``Frank copula'', ``Joe copula'', ``Clayton-Gumbel copula'', ``Joe-Gumbel copula'', ``Joe-Clayton copula'', ``Joe-Frank copula''\} together with corresponding rotations. No copula mentioned above reproduces the structure of the regression function in the resulting estimate.}
	
This observation can be explained by the fact
that none of the available parametric copula models for the vector $(Y,X_1)$ yields a non-monotone regression function
\begin{eqnarray} \label{A1}
m(x_1) = \mathbb{E}[Y | X_1 = x_1] &=& \int_{\- \infty}^{\infty} y c(F_Y(y), F_1(x_1)) d F_Y(y) = \int_0^1 F_Y^{-1}(y) c(y, F_1(x_1)) dy 
\end{eqnarray}

\begin{figure}[t]
\includegraphics[height=5.5cm,width=6cm]{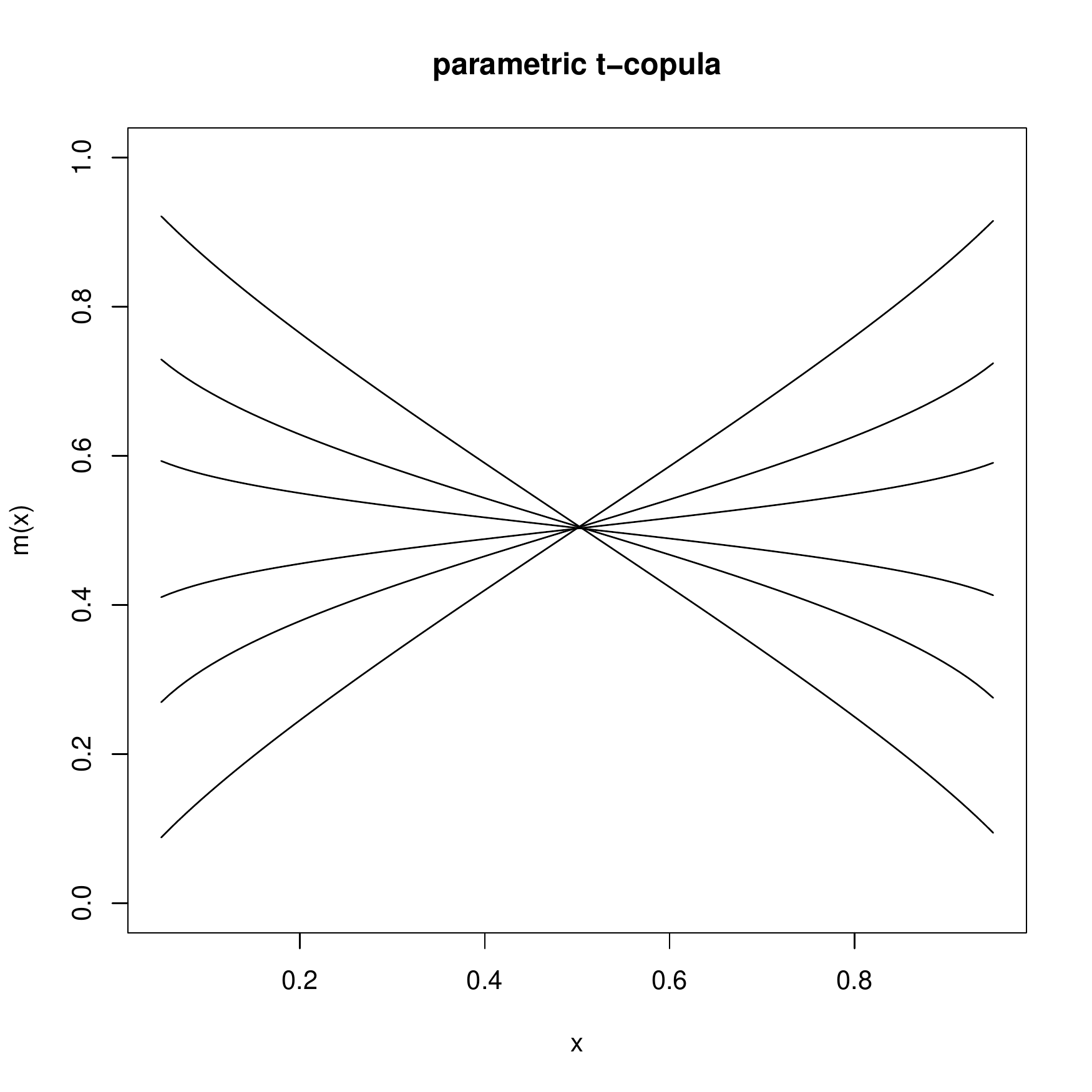}
\includegraphics[height=5.5cm,width=6cm]{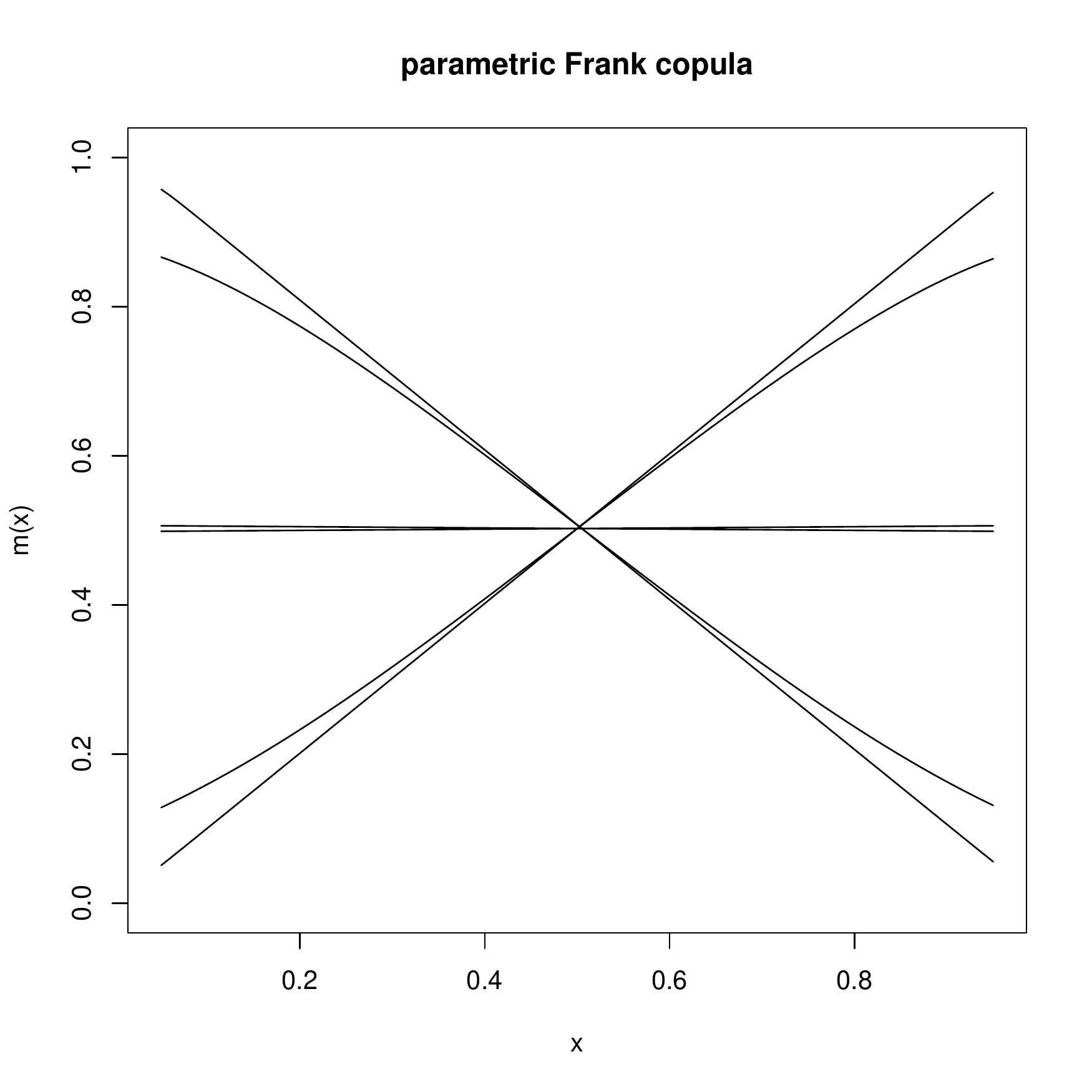}
\includegraphics[height=5.5cm,width=6cm]{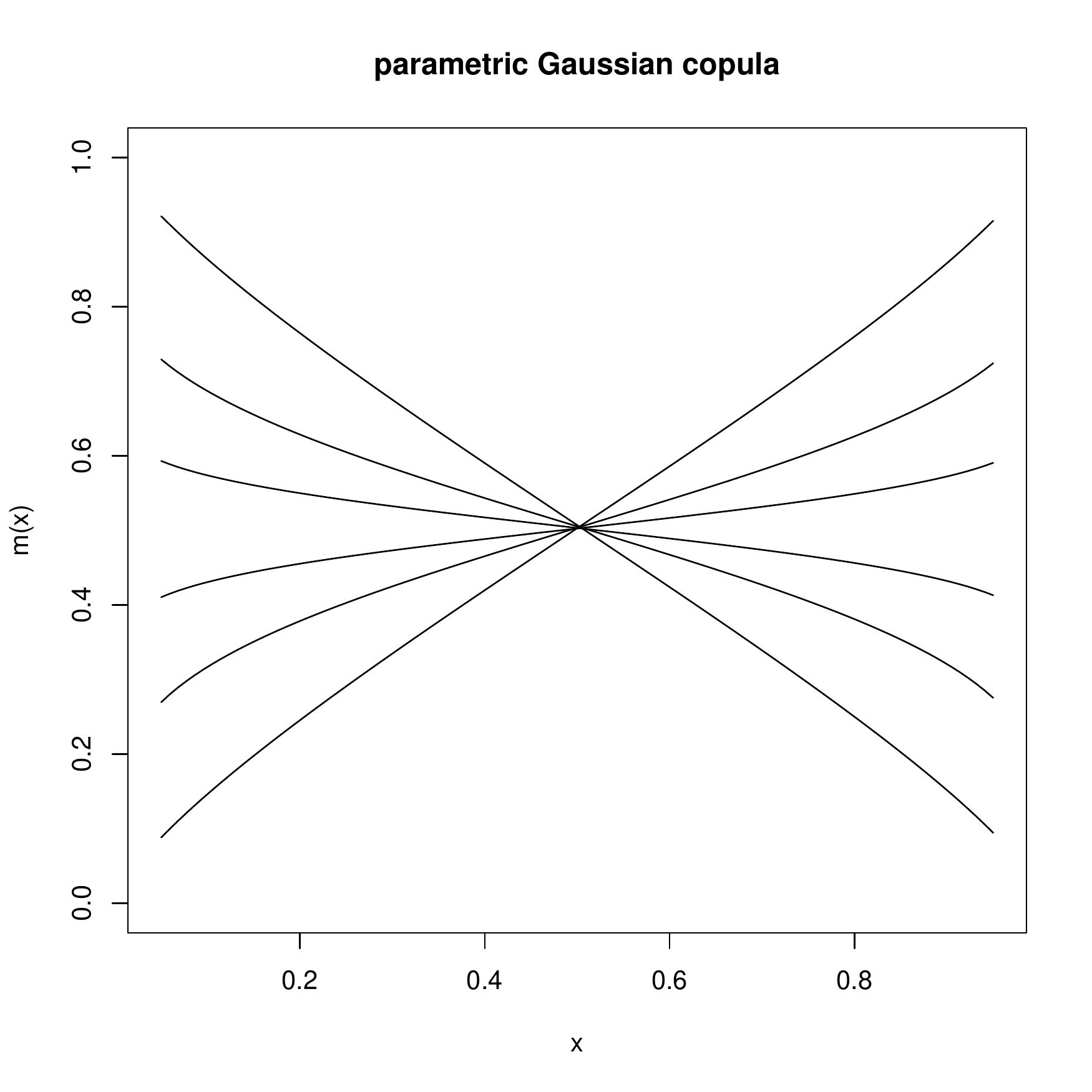}\\
\includegraphics[height=5.5cm,width=6cm]{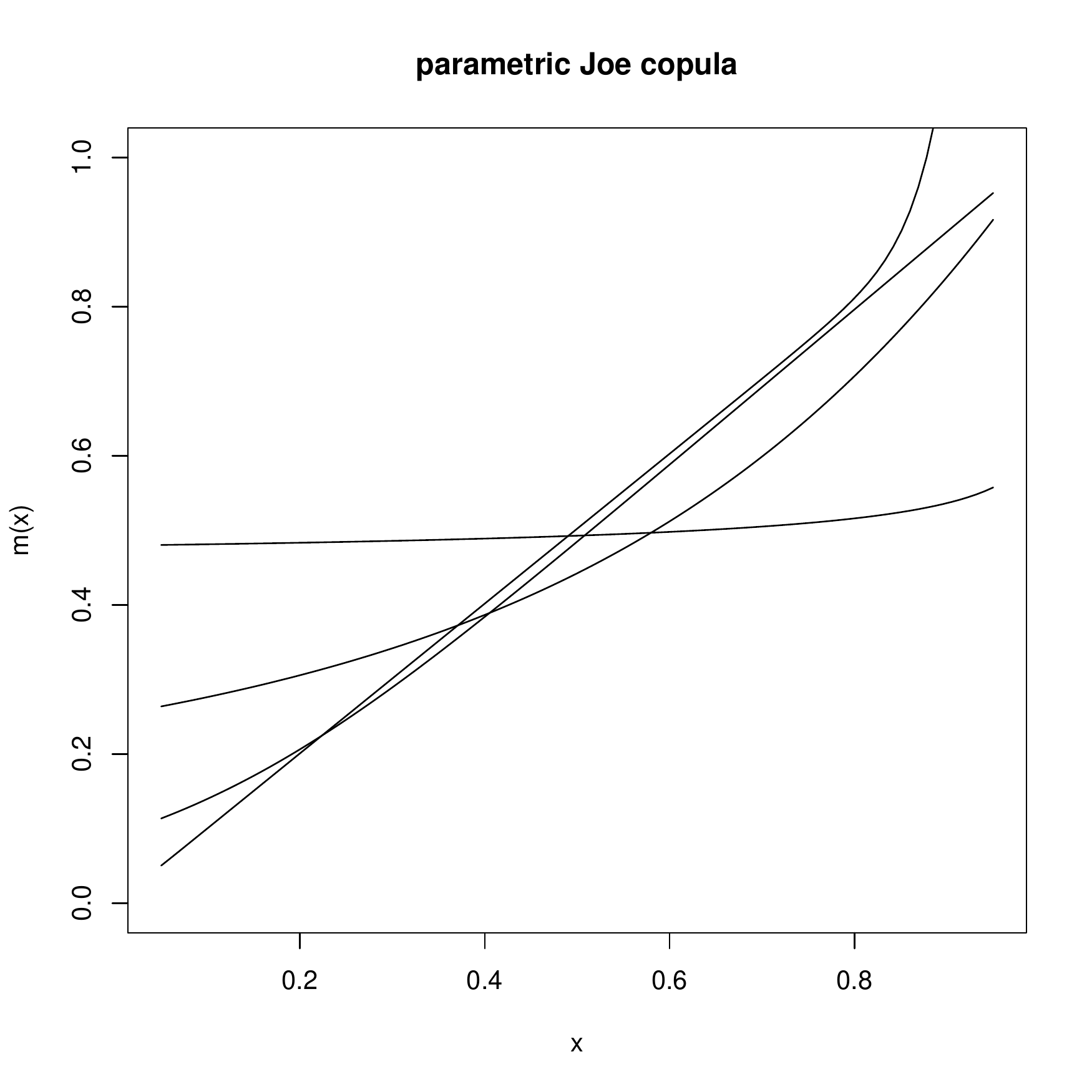}
\includegraphics[height=5.5cm,width=6cm]{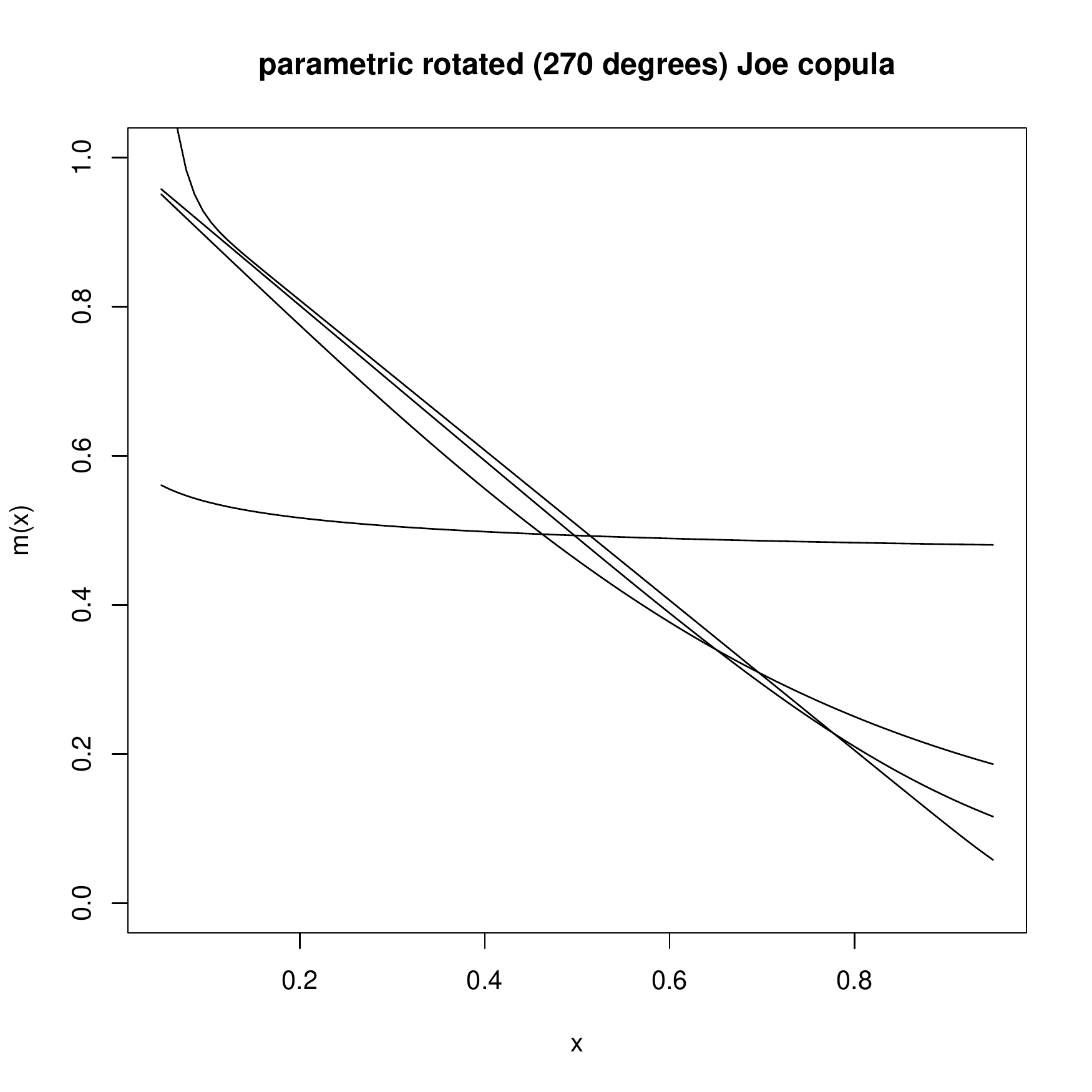}
\includegraphics[height=5.5cm,width=6cm]{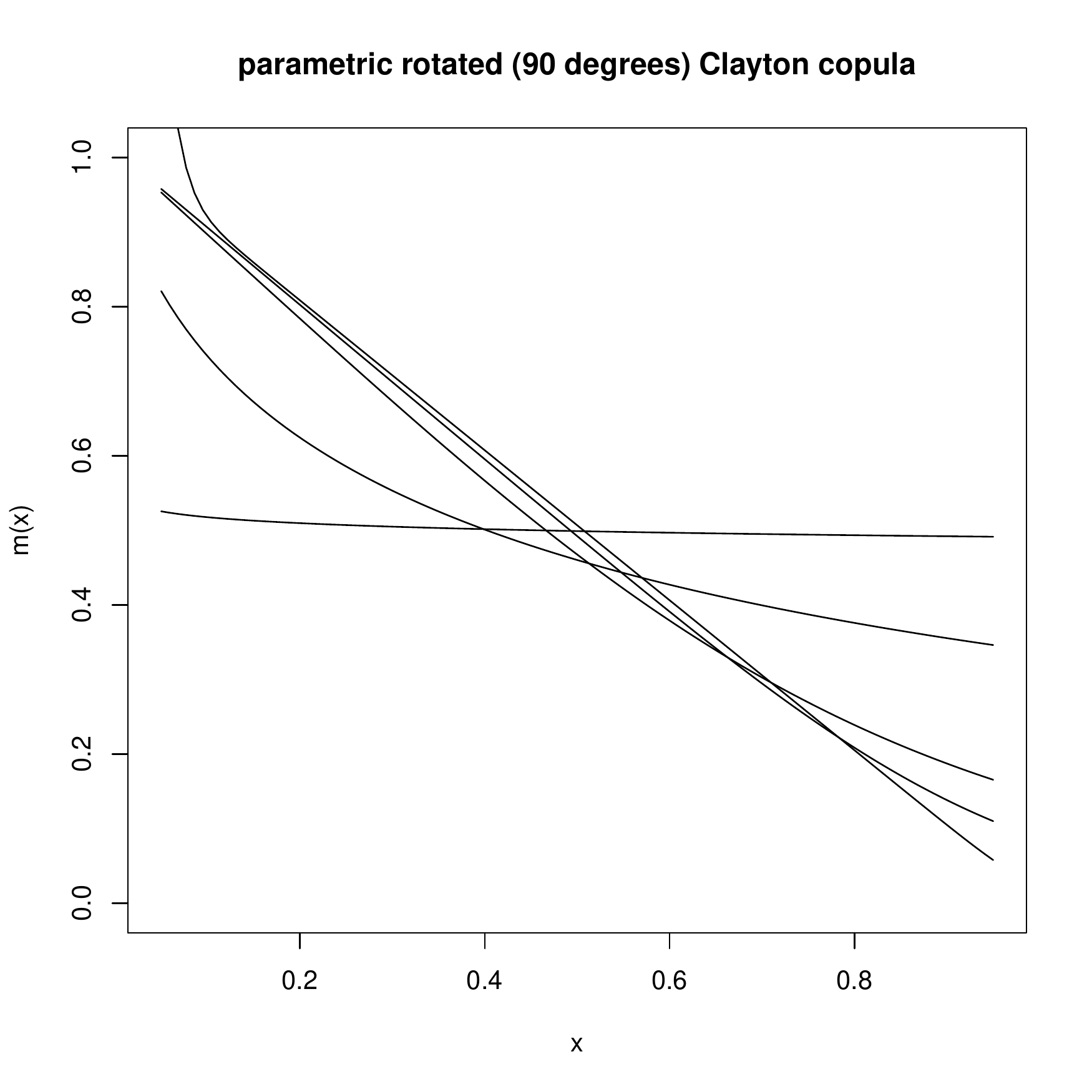}\\
\includegraphics[height=5.5cm,width=6cm]{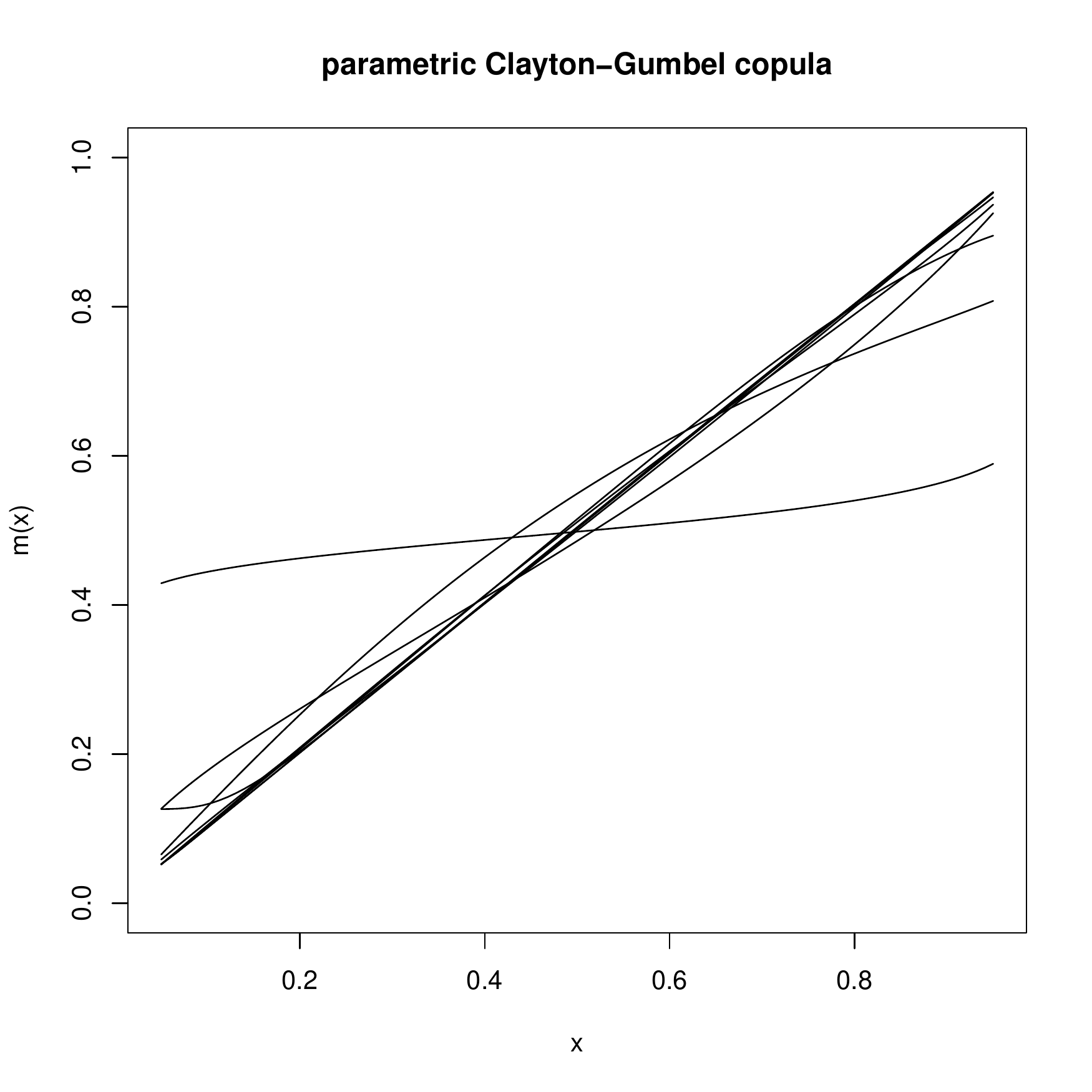}
\includegraphics[height=5.5cm,width=6cm]{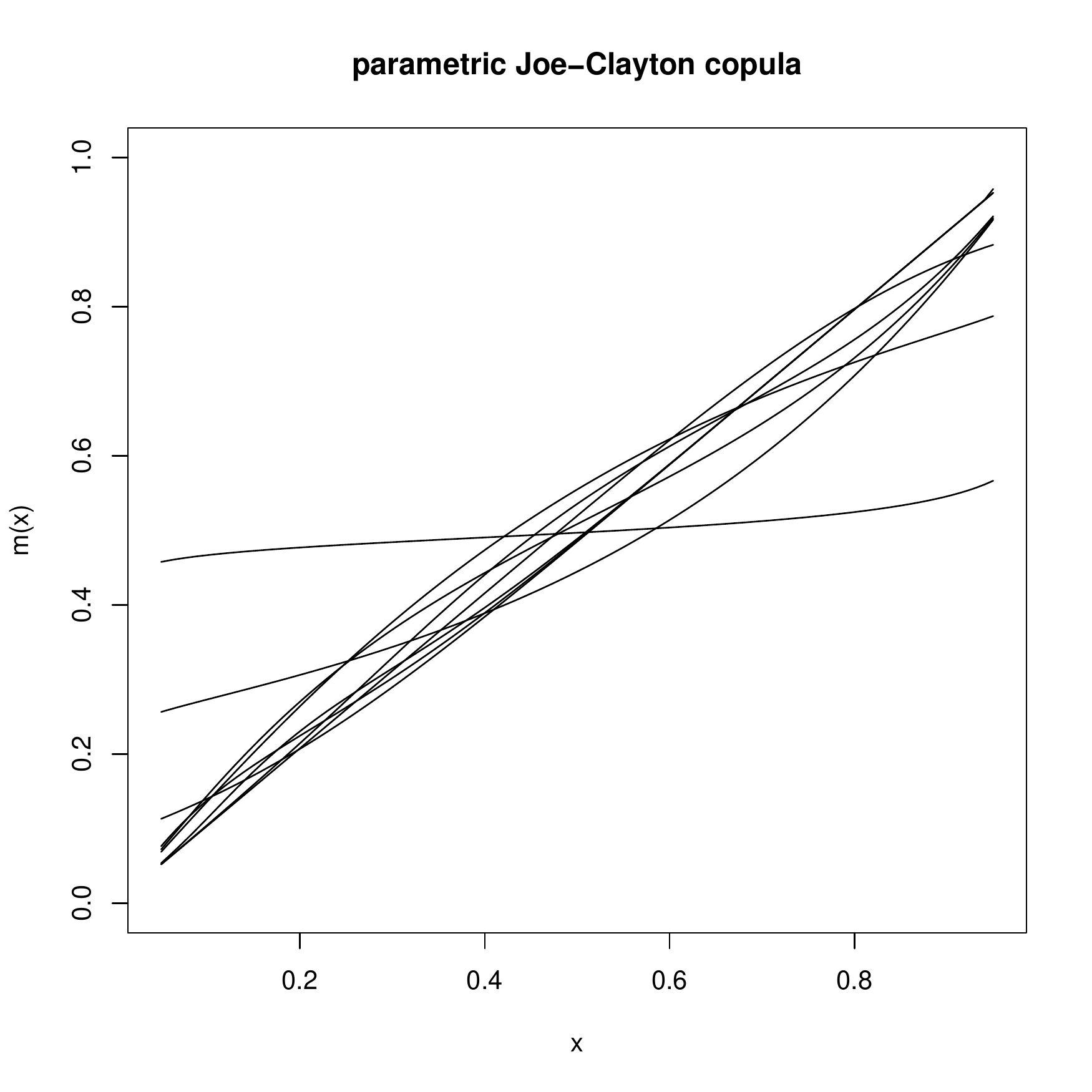}
\includegraphics[height=5.5cm,width=6cm]{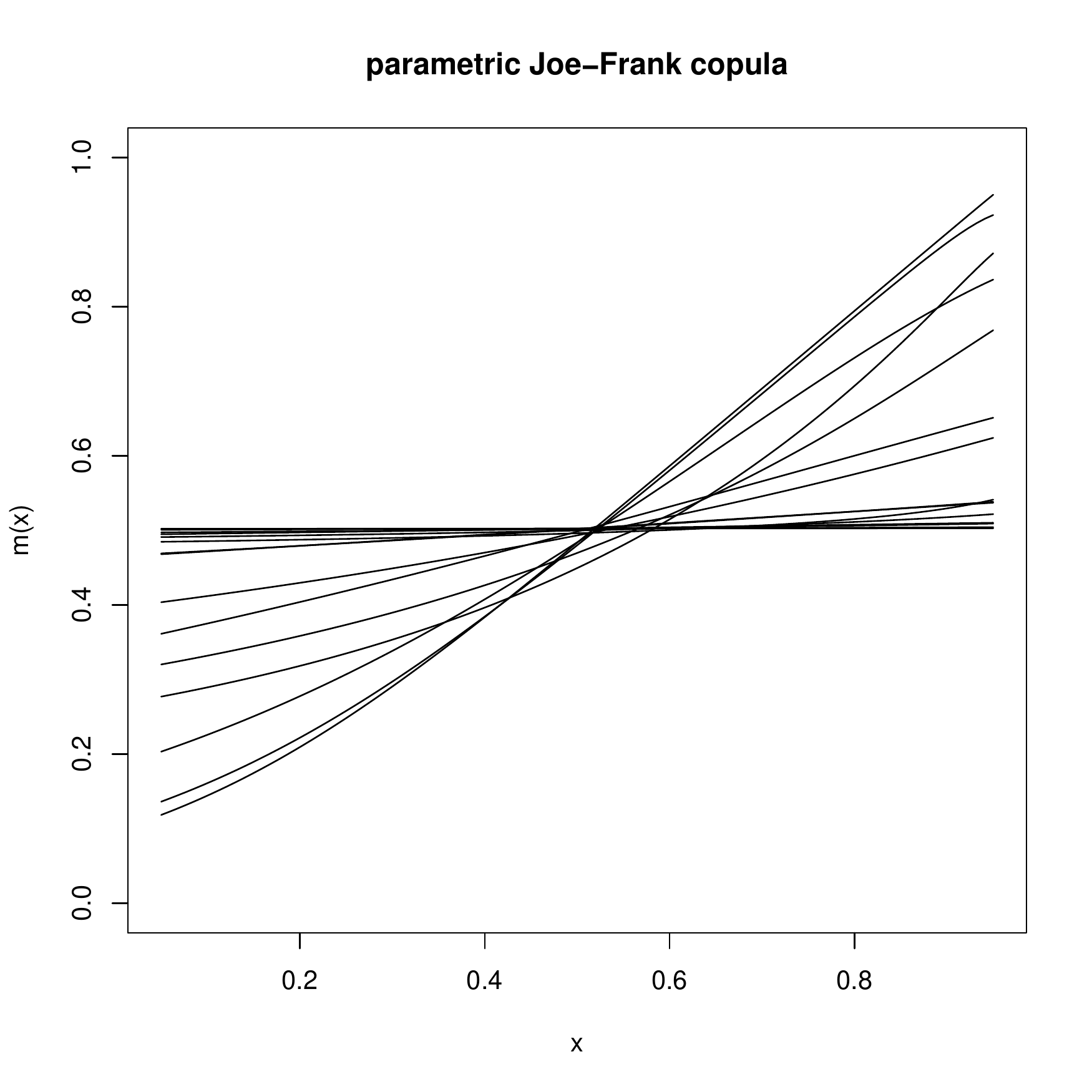}\\
\includegraphics[height=5.5cm,width=6cm]{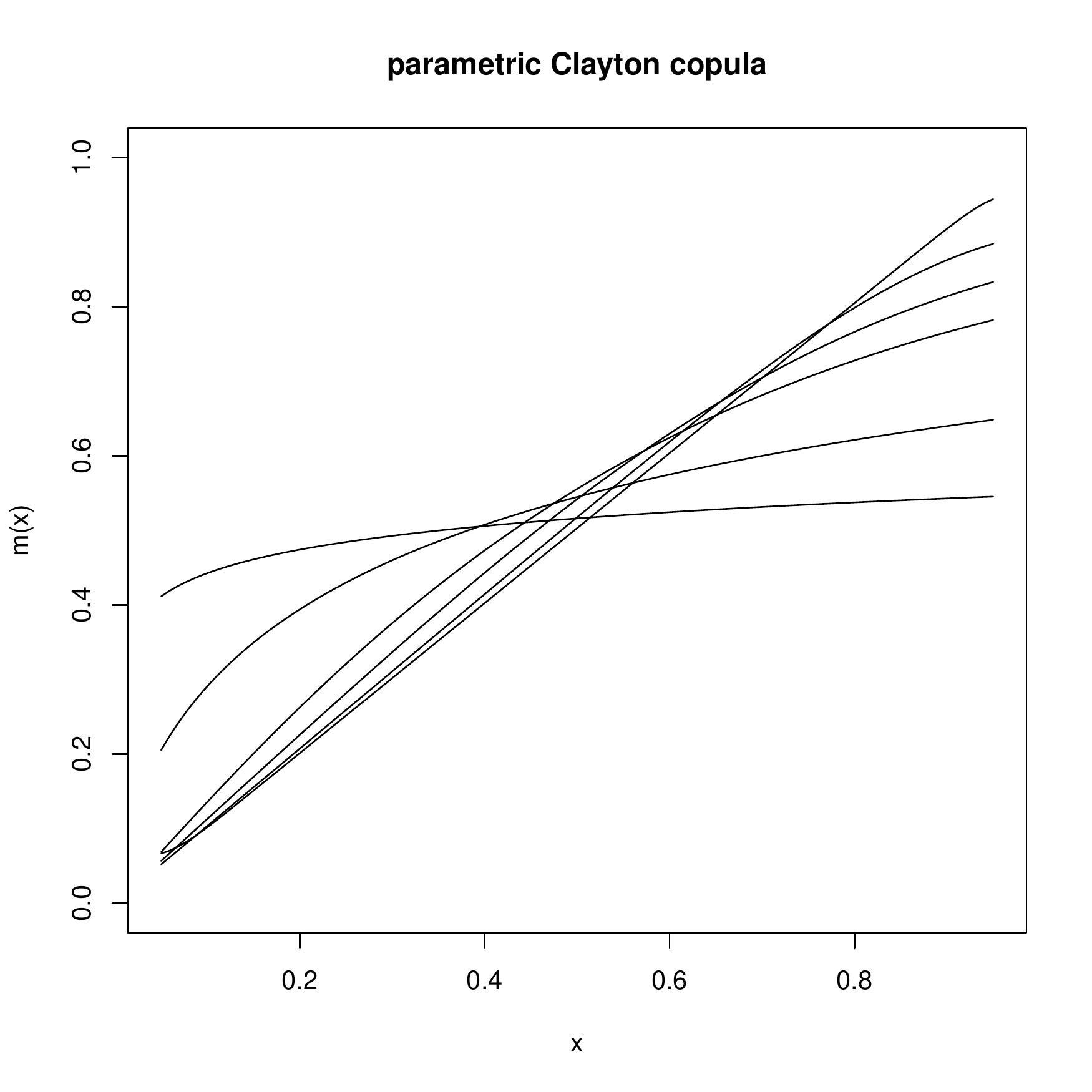}
\includegraphics[height=5.5cm,width=6cm]{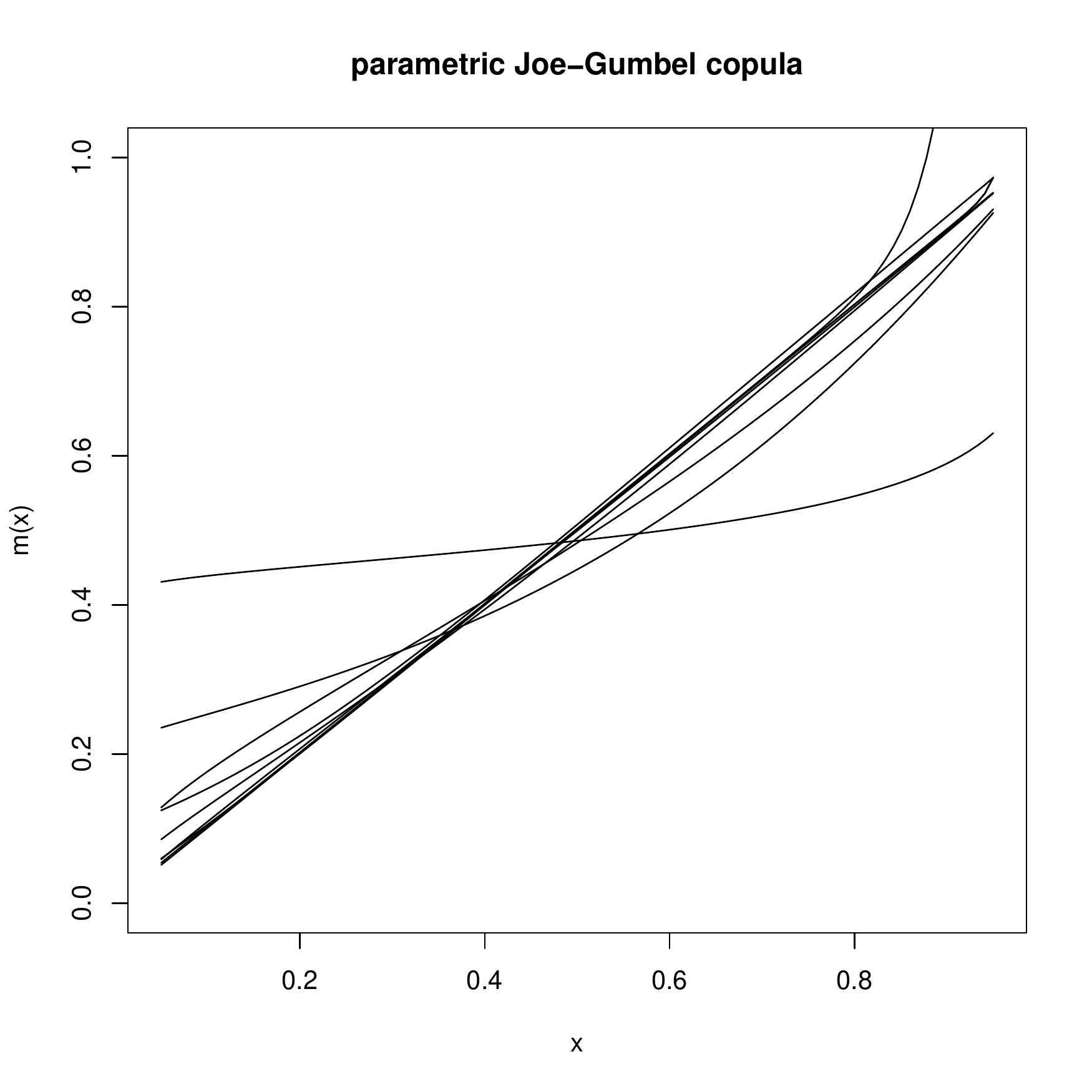}
\includegraphics[height=5.5cm,width=6cm]{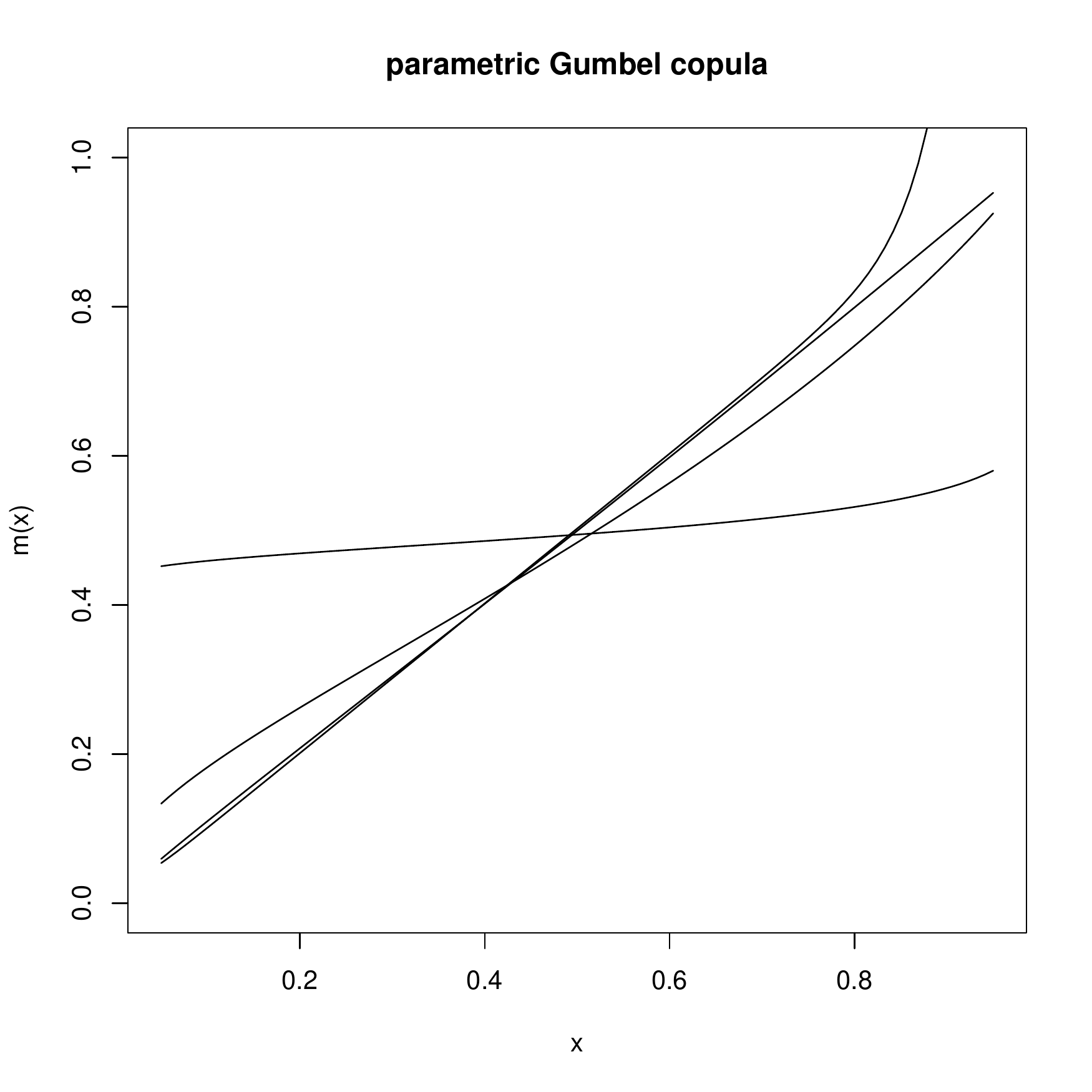}
\caption{\label{fig5}{\it The function (\ref{A1}) for commonly used parametric copula families (different parameters are used in each figure)}}
\end{figure}

In Figure \ref{fig5} this function is exemplarily displayed for various commonly used parametric copula models (and different parameters). Other results, which are not displayed here for the sake of brevity, show similar features. We observe that all of the commonly used parametric copula models lead to a monotone regression in \eqref{A1}. As a consequence we point out that model selection (for example by the AIC criterion) from a large class of commonly used parametric copula models will not improve the performance of the estimate. We conclude this section with two remarks

\begin{rem}\textcolor{white}{hello}
\begin{itemize} 
{\rm \item[(1)] Obviously, by definition of a copula, there exists a copula model corresponding to the model \eqref{reg} with $m(x_1)=(x_1-0.5)^2$, but on the basis of our numerical investigations this copula cannot be well approximated by any of the commonly proposed parametric copula models.
\item[(2)] The application of alternative estimates for the parameter of the copula does not lead to a significant improvement of the situation. For example investigations for the $L^2$-type estimator defined by
\begin{equation}
\hat{\param}_{L^2}=\underset{\param \in \Theta}{\argmin}\sum_{i=1}^n\left(Y_i-\hat{m}(X_i;\param_{L^2})\right)^2,
\end{equation}
with
\begin{equation}
\hat{m}(x;\param_{L^2})=\frac{1}{n}\sum_{i=1}^nY_ic(\hat{F}_Y(Y_i),\hat{F}_1(x);\param_{L^2})
\end{equation}
yield a picture very similar to the results presented here (these results are not displayed for the sake of brevity).  }
\end{itemize}
\end{rem}

\subsection{Two-dimensional predictors}

\begin{figure}
\includegraphics[height=6cm,width=6cm]{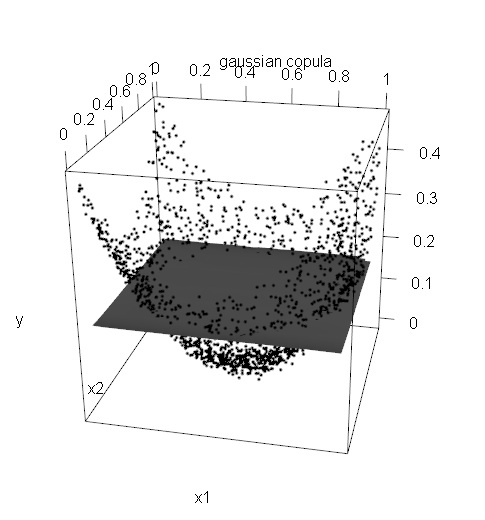}
\label{copplot_paraboloid_gauss}
\includegraphics[height=6cm,width=6cm]{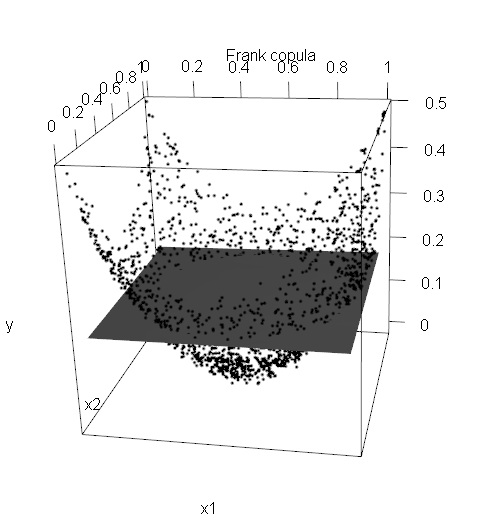}
\includegraphics[height=6cm,width=6cm]{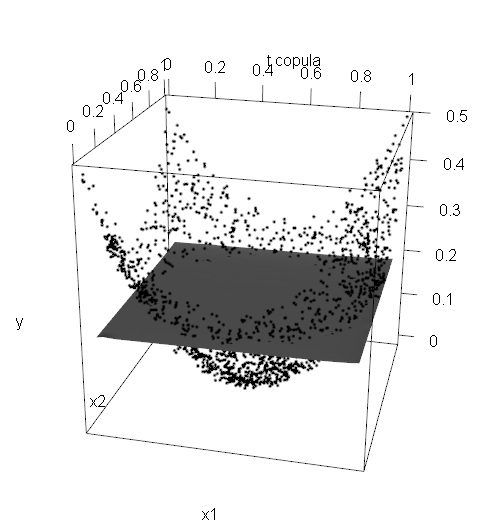}
\caption{\label{fig6} {\it Copula based regression estimates of the two-dimensional regression function \eqref{m3}. The copula in the estimate \eqref{estim} is chosen as
Gaussian copula (left panel), Frank copula (middle panel) and t-copula (right panel)}.}
\label{copplot_paraboloid}
\end{figure}

In this section we consider the case $d=2$. Obviously the observations of the previous paragraph carry over to higher dimensional  predictors if the regression is not monotone in one of the predictors and for this reason we exemplarily briefly consider the two-dimensional regression model
\begin{equation} \label{reg2}
Y_i=m(X_{i1},X_{i2}) +\sigma \varepsilon_i,~~i=1,\ldots , n,
\end{equation}
with   regression function
\begin{equation} \label{m3}
m(x_1,x_2)=(x_1-0.5)^2 + (x_2-0.5)^2,
\end{equation}
where the sample size is again $n=100$ and $X_{i1}$ and $X_{i2}$ are independent uniformly distributed on the interval $[0,1]$.
Some results for the Gaussian, Gumbel and t copula copula are displayed in Figure \ref{fig6},
and we observe the same problems as in the one-dimensional case. The considered parametric copula families are simply not flexible enough
such that the resulting estimate is able to reflect the curvature of the regression function.

\subsection{Vine copulae}\label{sec:vine} 
\begin{figure}[t]
\begin{center}
\includegraphics[height=6cm,width=6cm]{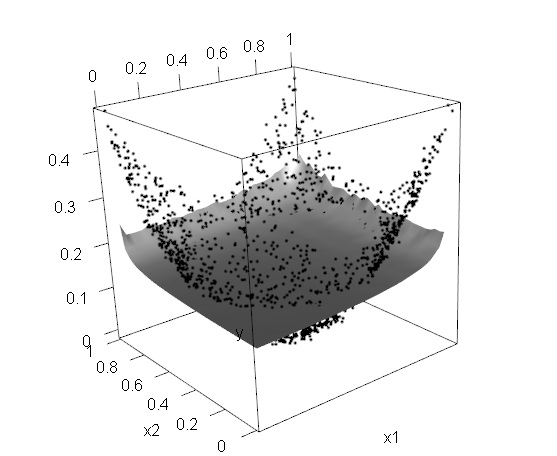}
\includegraphics[height=6cm,width=6cm]{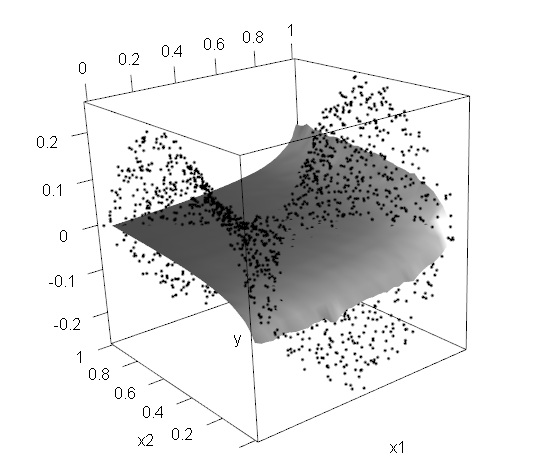}
\end{center}
\caption{\label{fig7} \it
Copula based regression estimates of the regression function in model \eqref{m3} (left panel), 
and \eqref{m5} (right panel). A vine copula selected by the AIC criterion has been used in the estimate \eqref{estim}. }
\end{figure}

In order to investigate more flexible classes of parametric copula models we briefly consider the concept of vine copulae, which is based on a  decomposition of the copula density into a product of bivariate copula densities according to a carefully chosen so called regular-vine structure.
The bivariate copula densities are then chosen from parametric  copula families by applying a model selection criterion. It has been argued by several authors  [see e.g. \cite{bedcoo2002,aas09}] that the resulting vine copula obtained by this pair-copula decomposition admits a flexible modelling of the dependency structure in the case of multiple covariates. In this section we investigate if this concept can be used to obtain improved copula based regression estimates. For a two dimensional predictor we obtain for the copula density of the random vector $(Y,X_1,X_2)$
\begin{eqnarray} \label{vine}
c(F_Y(y),F_1(x_1),F_2(x_2))=&c_{Y,X_1}(F_Y(y),F_1(x_1))c_{X_1,X_2}(F_1(x_1),F_2(x_2))\\
&\times c_{Y,X_2|X_1}(F_{Y|X_1}(y|x_1),F_{X_2|X_1}(x_2|x_1)), \nonumber
\end{eqnarray}
 where $c_{Y,X_1},c_{X_1,X_2}$ and $c_{Y,X_2|X_1}$ denote the copula densities associated with the distributions
  $\mathbb{P}^{Y_1,X_1},$ $\mathbb{P}^{X_1,X_2}$ and $\mathbb{P}^{Y,X_2|X_1}$ respectively. This decomposition is unique up to permutations of the variables.
 This pair-copula decomposition gives us the possibility to model the copula density $c$ in different ways by first selecting a R-Vine structure and then choosing the pair-copulae
 independently from a set of parametric copula families. In the implementation we used the $R$-package {\it VineCopula}
 and the copulae
 \{``independence copula'', ``Gaussian copula'', "t-copula", ``Clayton copula'', ``Gumbel copula'',
  ``Frank copula'', ``Joe copula'', ``Clayton-Gumbel copula'', ``Joe-Gumbel copula'', ``Joe-Clayton copula'', ``Joe-Frank copula''\} with corresponding rotations.
  The mean regression function $m$ is then defined by \eqref{estim} with copula \eqref{vine}, where the copulae for
  $c_{Y,X_1}$, $c_{X_1,X_2}$ and $c_{Y,X_2|X_1}$ are chosen by a function of the R package \textit{VineCopula} which determines a R-Vine tree-structure
  as well as the copula families with the corresponding estimated parameters using the Akaike information criterion [see \cite{dissmann11} for more details].
  In Figure \ref{fig7} we display a typical situation for model \eqref{reg2} with regression functions given in \eqref{m3} as well as,
\begin{eqnarray} 
&&m(x_1,x_2)=(x_1-0.5)^2 - (x_2-0.5)^2 \label{m5}
\end{eqnarray}
The sample size is again $n=100$ and the variance is $\sigma^2=0.01$. 
We observe that in all cases the copula based regression method does not yield estimates which reflect the qualitative behaviour of the regression function.
These results show that even the rather large family of vine-copulae is not flexible enough to adapt for the unknown copula structures imposed
by the models \eqref{m3} and \eqref{m5}.

\section{Conclusions} \label{sec4}

\begin{figure}[t]
\includegraphics[height=6cm,width=6cm]{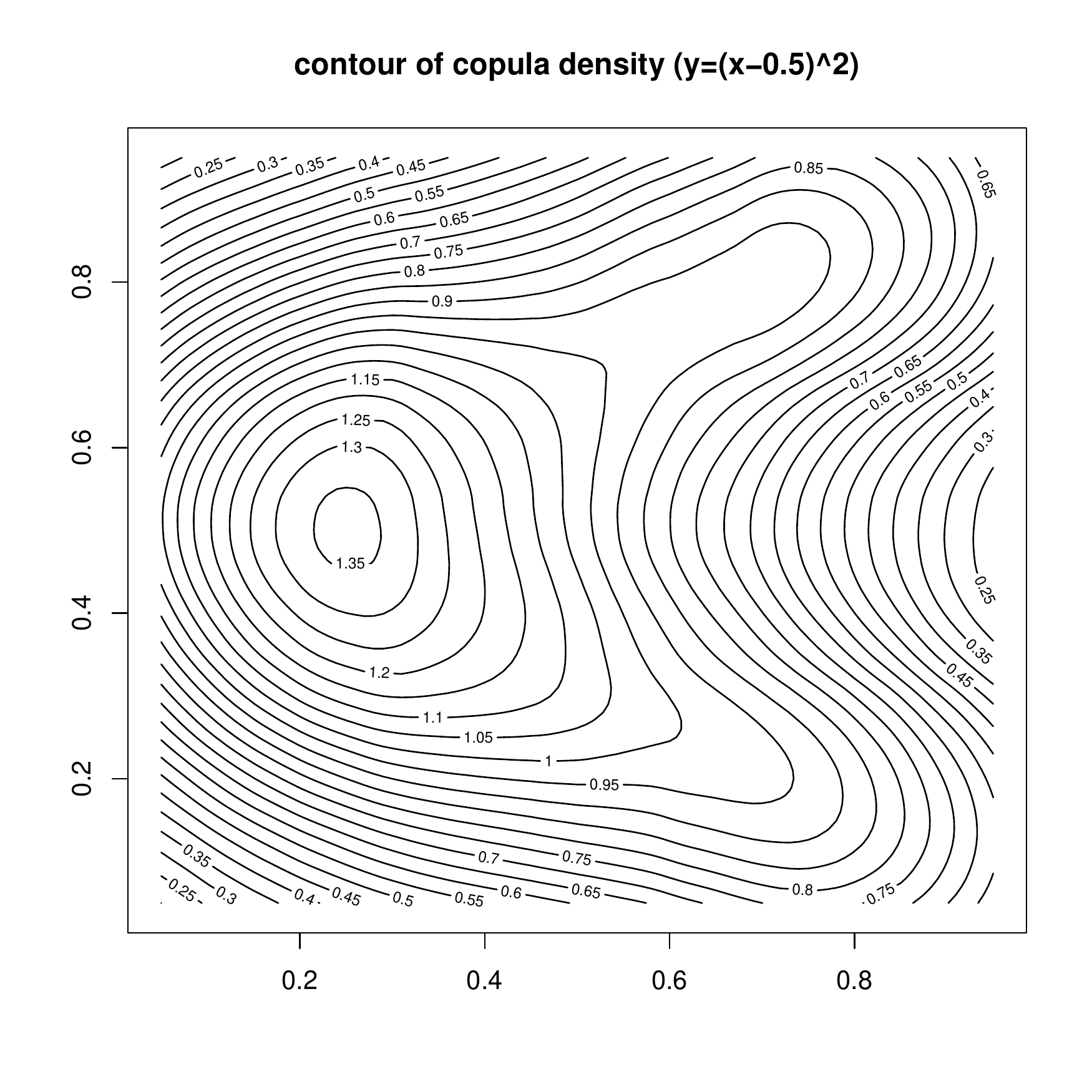}
\includegraphics[height=6cm,width=6cm]{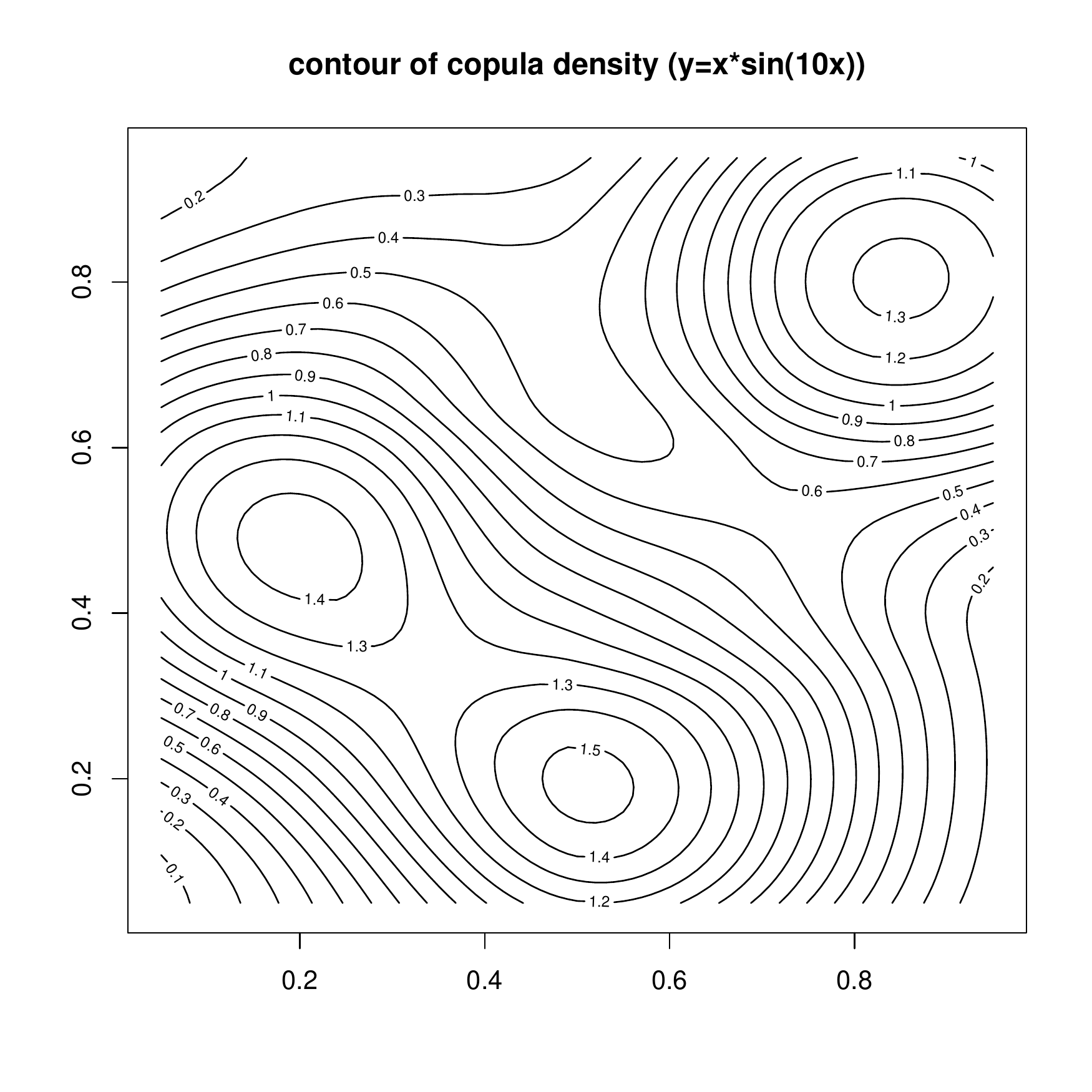}
\includegraphics[height=6cm,width=6cm]{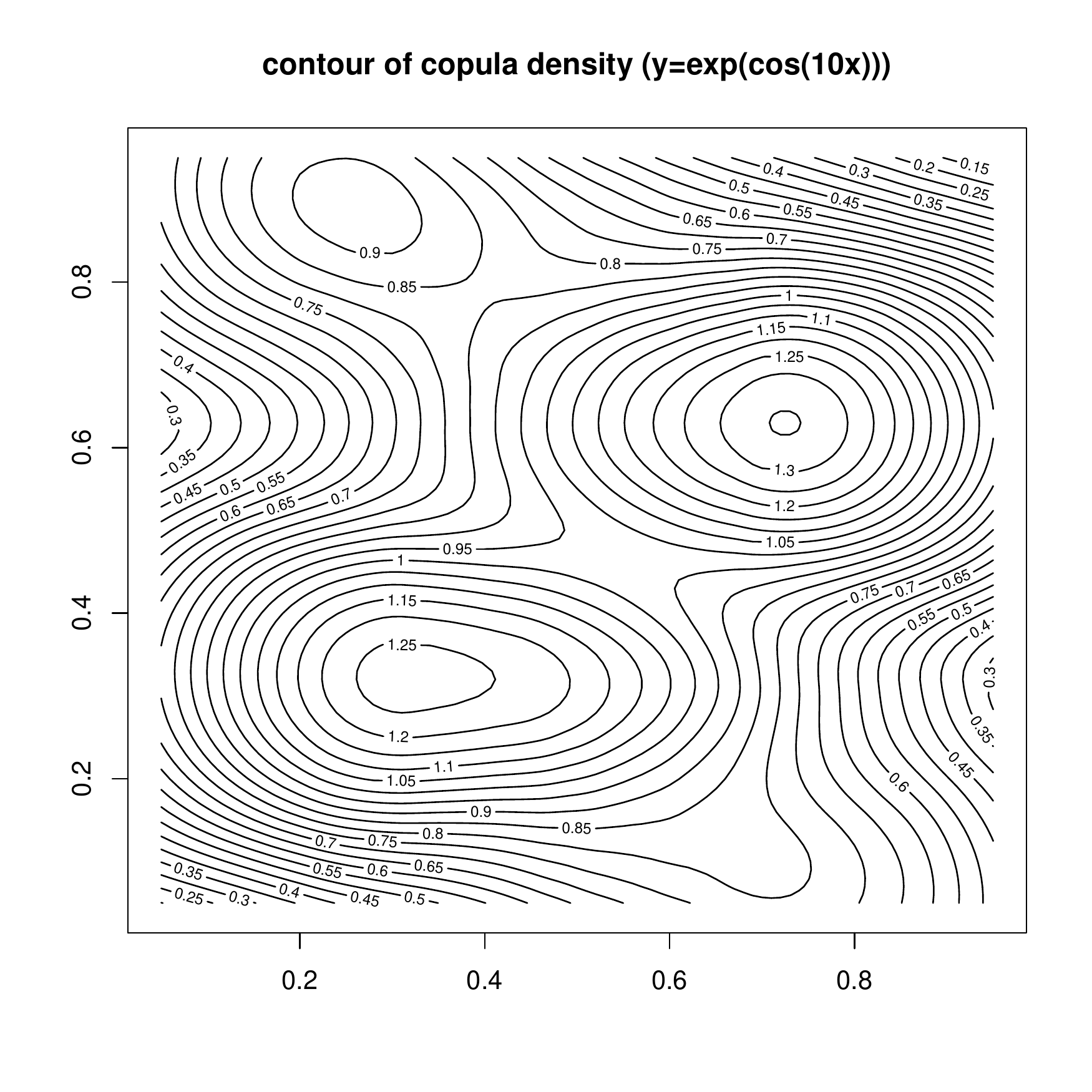}
\caption{\label{fig8} \it Simulated contour plots of the copula density corresponding to the one-dimensional regression model \eqref{m2} (left panel), the function $m(x)=x \sin(10x)$ (middle panel) and the function $m(x)=\exp (\cos(10x))$ (right panel) with Gaussian errors.
}
\end{figure}

{In this note we have studied some properties of a semiparametric copula-based regression estimate which has been recently proposed in \cite{noh13}, and, more broadly, the types of regression dependence that can be obtained from commonly used copula families. Our simulations confirm that the approach of \cite{noh13} is attractive if the dependency structure of the data can be specified accurately.} On the other hand -- if the true copula structure has been misspecified -- the approach { often }does not yield reliable estimates of the regression function. The reason for these problems consists in the fact that all commonly used parametric copula families produce a regression $ m({\bf x})=\mathbb{E}[Y | {\bf X}={\bf x}]$ which is monotone in any of the components of the explanatory variable $\X$. As a consequence non-monotone features of the regression function cannot be reproduced by the copula-based regression estimate. In Figure \ref{fig8} we display level sets of a copula density corresponding to some non-monotone regression functions. We observe that these differ substantially from the sets of all parametric copula densities. Future research is necessary to develop more flexible parametric copula models reflecting these structures. If this is possible we expect that the method proposed by \cite{noh13} will be able to reproduce non-monotone features of the regression function.

\bigskip
\bigskip




\bigskip
\bigskip

{\bf Acknowledgements.} The authors would like to thank Martina Stein, who typed parts of this manuscript with considerable technical expertise.
This work has been supported in part by the Collaborative
Research Center ``Statistical modeling of nonlinear dynamic processes'' (SFB 823, Teilprojekt A1, C1, A7) of the German Research Foundation (DFG). We are also grateful to C. Czado for interesting discussion about the concept of vine copulae {and to Axel B\"ucher and Anouar El Ghouch for comments on a preliminary version of this manuscript}.

\bibliographystyle{apalike}
\bibliography{copula_regression}

\end{document}